\documentclass[12pt,preprint,pre,epsfig,address]{article}
\usepackage[pdftex]{graphicx}
\usepackage{subcaption}

\usepackage[english]{babel}
\usepackage{amsmath, bm}
\usepackage{amssymb}
\usepackage{afterpage}
\usepackage{dsfont} 
\usepackage{braket}
\usepackage[colorlinks = true,
linkcolor = blue,
urlcolor  = blue,
citecolor = purple,
anchorcolor = blue]{hyperref}

\usepackage{graphicx}

\usepackage[dvipsnames]{xcolor}
\usepackage{color}

\newcommand{\be}{\begin{equation}}
\newcommand{\ee}{\end{equation}}
\newcommand{\bel}[1]{\begin{equation}\label{#1}}
\newcommand{\bea}{\begin{eqnarray}}
\newcommand{\eea}{\end{eqnarray}}
\newcommand{\balign}{\begin{align}}
\newcommand{\ealign}{\end{align}}
\newcommand{\ba}{\begin{array}}
\newcommand{\ea}{\end{array}}
\newcommand{\bfig}{\begin{figure}}
\newcommand{\efig}{\end{figure}}

\newcommand{\nn}{\nonumber}

\allowdisplaybreaks

\begin{document}

\title{Some connections between the Classical Calogero-Moser model and the Log Gas}
\author{Sanaa Agarwal$^{1,2}$, Manas Kulkarni$^{2}$ and Abhishek Dhar$^{2}$}

\maketitle

{\small
$^{1}$ Birla Institute of Technology and Science, Pilani - 333031, India

$^{2}$ International centre for theoretical sciences, Tata Institute of Fundamental Research,

\hspace{4.5pt} Bangalore - 560089, India
\\

}

\begin{abstract}
In this work we discuss  connections between a one-dimensional system of $N$ particles interacting with a repulsive inverse square potential and  confined in a harmonic potential (Calogero-Moser model) and the  log-gas model which appears in random matrix theory. 
Both models have the same minimum energy configuration, with the particle positions given by the zeros of the Hermite polynomial. Moreover, the Hessian describing small oscillations around equilibrium are also related for the two models. The Hessian matrix of the Calogero-Moser model is the square of that of the log-gas. We explore this connection further by studying finite temperature equilibrium properties of the two models through Monte-Carlo simulations.  
In particular, we study the single particle distribution and the marginal distribution of the boundary particle which, for the log-gas, are respectively given by the Wigner semi-circle and the Tracy-Widom distribution. For particles in the bulk, where typical fluctuations are Gaussian, we find that numerical results obtained  from small oscillation theory are in very good agreement with the Monte-Carlo simulation results for both the models. For the log-gas, our findings  agree with  rigorous  results from random matrix theory.
\end{abstract}
\newpage
\tableofcontents
\newpage
\section{Introduction}
The study of systems of classical interacting particles in one-dimension (1D) with Hamiltonian dynamics is of interest from many points of view. From a dynamical point of view such systems are broadly classified as integrable and non-integrable and they  are known to usually show qualitatively different behaviour when one looks at properties such as dynamical correlation functions, transport and equilibration. On the other hand as far as equilibrium properties are concerned one does not expect qualitative differences in behaviour arising out of integrability or otherwise of the Hamiltonian. 
Such finite temperature equilibrium properties of classical systems remains a subject of great interest till date. Seemingly unrelated systems often have common statistical properties and identifying universal features is one of the most interesting questions. Two very well-known one-dimensional models of interacting many particle systems are the so called log-gas (LG) and the Calogero-Moser (CM) model. The main aim of the present paper is a detailed study of the finite-temperature equilibrium properties of these two systems, pointing out surprising and unexpected connections between the two systems.

The log-gas model~\cite{forrester2010log,nagao1995asymptotic} represents particles in 1D interacting with each other via a repulsive logarithmic potential and confined to move in a harmonic confining potential.  The log-gas has a  strong connection to random matrix theory~\cite{bourgade2014bulk,erdos2012universality,forrester2010log}. 
The joint probability distribution function (pdf) of the eigenvalues of a $N \times N$ random matrix  is given precisely by the equilibrium Boltzmann distribution of the positions of the $N$-particle log-gas~\cite{ameur2011fluctuations,forrester2010log} --- with the three important symmetry classes orthogonal, unitary and  symplectic, of random matrices corresponding to $\beta =1,2,4$ in the log-gas distribution. Using this connection it has been possible to obtain many rigorous results for equilibrium properties of the log-gas at special values of the inverse temperature ($T^{-1}=\beta=1,2,4$)~\cite{deift2006universality,erdos2012universality,forrester2010log} while  less is understood for other values of $\beta$~\cite{forrester2010log}. Some well known results include the single-particle distribution which is given by the Wigner semi-circle law and the marginal distribution of the edge particle given by the  Tracy-Widom (TW) form~\cite{tracy1998correlation,tracy2009distributions,widom1999relation,baik2005phase,baker1997finite,nagao1995asymptotic,nagao1998transitive}. Other results for the log-gas following from the random matrix connection include computation of density-density correlations~\cite{forrester2010log,mehta2004random},   and more recently  it has been shown that the marginal distributions of $k << N$ well-separated ordered bulk  particles is Gaussian~\cite{gustavsson2005gaussian,o2010gaussian,zhang2015gaussian}. 
Somewhat surprisingly, although there have been several numerical studies~\cite{bornemann2009numerical} of Gaussian random matrix ensembles, the numerical simulation of the log-gas and numerical verification of various theoretical predictions are quite rare in the existing literature and in this paper we fill this gap.

On the other hand, the Calogero-Moser model~\cite{calogero1975exactly,calogero1969_1,calogero1971solution, moser1976three} describes a Hamiltonian system with particles interacting via a repulsive inverse square potential and confined in a harmonic potential. It is one of the classic examples of an  integrable system, and apart from being an important solvable model in both classical and quantum physics, it has become ubiquitous in  areas ranging from  soliton physics, string theory, condensed matter physics, random matrix theory  \cite{bogomolny2009random,kulkarni2017emergence,polychronakos2006physics} as well as in  pure mathematics. 
For the CM model, the integrals of motion can be constructed from Lax pairs \cite{olshanetsky1981classical,perelomov1990integrable}. The  model admits multi-soliton solutions and is known to display an interesting duality \cite{abanov2011soliton}. Although much has been studied in terms of collective field theory formalism \cite{aniceto2010poisson, stone}, nonlinear dynamics, solitons \cite{abanov2011soliton}, quenches \cite{franchini2015universal,franchini2016hydrodynamics} in the above model, the finite temperature equilibrium properties of the CM system 
 are less well known. Connections with random matrix theory have also been noted. In \cite{bogomolny2009random} the Calogero-Moser potential (with harmonic pinning) was found to describe the eigenvalue distribution of a  random matrix ensemble  constructed from the Lax matrix of the Calogero-Moser Hamiltonian (without  harmonic pinning potential). The level spacing distribution was found to be described by a form analogous to the Wigner surmise, but with very strong level repulsion. However other properties of the many-particle distribution were not investigated.  In fact in contrast to the log-gas,  there are few results on the equilibrium physics of the Calogero-Moser system. 

At first sight the log-gas system and Calogero-Moser system appear to be quite unrelated but it turns out that unexpected connections between the two models have been noted. For the classical case, it was noted by Calogero \cite{calogero1977equilibrium} that the positional configuration of particles corresponding to the minimum of the potential is identical for both models and is given by the zeros of the Hermite polynomial. Furthermore it was pointed out that the Hessian matrices, corresponding to small oscillations around the minimum, are also related for the two models. In the quantum case, it was shown that the probability distribution corresponding to the ground state wave function of the Calogero-Moser model has a Boltzmann form with the potential  being precisely that of the log-gas model~\cite{sutherland1971quantum,sutherland1972exact}.

Given the several unexpected connections between the two models it is natural to explore this further in the context of classical equilibrium physics and this is the main aim of the present paper. An example of an interesting question that one could ask is regarding the distribution of the edge particle which is known to be of the Tracy-Widom form for the log-gas.  In a recent paper the question of universality of this result for general 1D interacting systems was investigated and it was shown that the form is in fact very different for a 1D coulomb gas (interaction potential between particles given by modulus of distance) \cite{dhar2017}. In the present paper we ask the same question for the Calogero-Moser model. More generally we explore other possible connections between the two models through a detailed study of their equilibrium correlations and various one-point distribution functions. We present results from extensive Monte-Carlo simulations of the two models as well as numerical computations from the Hessian matrices  corresponding to small oscillations.

We  summarize here our findings. We consider a system of $N$ particles with positions $(x_1,x_2,\ldots,x_N)$ that are in thermal equilibrium at temperature $T=1/\beta$  in a many-body potential described by either the LG or the CM models. Our main results are: 

\noindent (i) We verify that the single particle density distribution of the particles in both models converges in the continuum limit, as $N \to \infty$, to the Wigner semi-circle form at all finite temperatures, with differences from the log-gas in the  finite-size effects.  

\noindent(ii) For the CM model, we provide numerical evidence that the fluctuations of the edge particle $\delta x_N = x_N- \langle x_N \rangle$ scale as $ \langle \delta x_N^2 \rangle \sim N^{-2/3}$ in contrast to the well-known scaling form  $ \langle \delta x_N^2 \rangle \sim N^{-1/3}$ for the log-gas. We find that the scaling function describing the typical fluctuations is non-Gaussian but different  from the Tracy-Widom  for  $\beta=1,2,4$.

\noindent(iii) We compute marginal distributions of bulk particles $x_k$ with $1<<k<<N$ and find that they are Gaussian distributed with variances $\langle \delta x_k^2 \rangle \sim   1/N$ for the CM model and $\langle \delta x_k^2 \rangle \sim   \log (N)/N$ for the LG model. For the LG model, we compare our results with the analytical predictions in \cite{gustavsson2005gaussian,o2010gaussian,zhang2015gaussian}.

\noindent(iv) We compute two-point correlations $\langle \delta x_i \delta x_k \rangle$ and verify the scaling form $\langle \delta x_i \delta x_k \rangle=(1/N)~f(i/N,k/N)$ where $f$ is a scaling function. This is valid except for $|i-k|<< N$, with different scaling functions for the LG and CM system. 

\noindent(v) In all cases, we find that two-point correlations obtained from our Monte-Carlo  simulations are in close agreement with the correlations obtained from the inverse of the Hessian matrices.

Our paper is organized as follows. In Sec~\ref{mod}, we define the two models,
namely the log-gas and the Calogero-Moser systems and state some known results on their minimum energy configurations and the Hessian matrices corresponding to small oscillations around the minima.
In Sec.~\ref{eqb}, we present numerical results from Monte-Carlo simulations on various probability distributions and two-point correlation functions. The results on correlation functions are compared with earlier known rigorous results (for the log-gas) and with those obtained from inverse of the Hessian matrix (for both systems). 
We summarise our results along with an outlook in Sec.~\ref{conc}. Some of the  numerical evidence demonstrating equilibration of the Monte-Carlo simulations are 
given in the appendix.

\section{Model and summary of some known results}
\label{mod}
Here we first define the  log-gas and the Calogero-Moser Hamiltonians. We consider a system of $N$ particles with positions $(x_1,x_2,\ldots,x_N)$ and momenta $(p_1,p_2,\ldots,p_N)$. 
The log-gas system is given by the following Hamiltonian:
\begin{align}
    H_{\rm log} &= \sum_{i=1}^N\frac{1}{2m}p_i^2 +  V_{\rm log}(x_1,x_2,\ldots,x_N), \nn \\
{\rm where}~~V_{\rm log} &= \sum_{i=1}^N\frac{1}{2}m w^2 x_i^2 - \sum_{i=1}^N\sum_{\substack{j=1\\j\neq i}}^N\frac{1}{2} \log|x_i - x_j|~.
    \label{log_gas_ham}
\end{align}
This describes particles in 1D interacting via a logarithmic  interaction potential and confined in a harmonic trap. 
 For the Calogero-Moser system in a harmonic potential, the Hamiltonian is given by:
\begin{align}
    H_{\rm cal} &= \sum_{i=1}^N\frac{1}{2m}p_i^2 + V_{\rm cal}(x_1,x_2,\ldots,x_N) \nn \\
{\rm where}~~V_{\rm cal}&=   \sum_{i=1}^N\frac{1}{2}m w^2 x_i^2 + \sum_{i=1}^N\sum_{\substack{j=1\\j\neq i}}^N\frac{1}{2} \frac{g^2}{(x_i - x_j)^2}~.
    \label{cal_ham}
\end{align}
This describes particles in 1D interacting via an inverse square interaction confined in a Harmonic trap. For studies of the equilibrium physics, the parameters $m,w,g$ can be scaled out and for the rest of the 
 paper, we set $m = g = w = 1$. Setting also  Boltzmann's constant  $k_B$ = 1 
the only relevant parameter is the inverse temperature  $\beta= {1}/{T}$.

\subsection{Minimum energy configuration}

The minimum energy configuration for the log-gas is given by the condition that the force $F_i$ on each particle vanishes: 
\begin{equation}
F_i= -{\partial V_{\rm log}}/{\partial x_i} = -x_i + \sum_{\substack{j=1\\j\neq i}}^{N} (x_i - x_j)^{-1}=0, ~~~{\rm for}~i=1,2\ldots,N~.
    \label{min_energy_log}
\end{equation}
It is well known that the solution for these set of equations is given by zeros of the Hermite polynomials  $H_N(y)$ \cite{mehta2004random, doi:10.1137/0517035}, which we denote by $z^{(N)}_i$, for $i=1,2,\ldots N$.
On the other hand the minimum energy configuration for Calogero-Moser model is given by the equation
\begin{equation}
{\partial V_{\rm cal}}/{\partial x_i} =  x_i - 2\sum_{\substack{j=1\\j\neq i}}^{N} (x_i - x_j)^{-3} = 0,~~~{\rm for}~i=1,2\ldots,N~.
    \label{min_energy_cal}
\end{equation}
As was shown first by Calogero~\cite{calogero1981matrices}, it turns out that this equation is also satisfied by the Hermite zeros. Here, we present a simpler proof of this result.

We first note that the CM potential can be expressed in the form
\begin{equation}
    V_{\rm cal} =  \sum_{i=1}^N F_i^2/2+N/2~,
    \label{vcal_equals_Flog}
\end{equation}
with $F_i$ given in Eq.~\ref{min_energy_log}.  This follows since
\begin{align}
    \sum_{i = 1}^N F_i^2 &= \sum_{i = 1}^N \left[ x_i^2 + \sum_{\substack{j = 1\\j\neq i }}^N \frac{1}{(x_i - x_j)^2} \right] +\sum_{\substack{i,j,k = 1\\i\neq j \neq k}}^N  \frac{1}{(x_i - x_j)(x_i - x_k)} - 2 \sum_{\substack{i,j = 1\\i\neq j }}^N  \frac{x_i}{(x_i - x_j)}~. \label{fsq} 
\end{align}
The second term above can be symmetrized to give
\begin{align}
\label{fid}
    \sum_{\substack{i,j,k = 1\\i\neq j \neq k}}^N  \frac{1}{(x_i - x_j)(x_i - x_k)} &= 
\frac{1}{3} \left[  \frac{1}{(x_i - x_j)(x_i - x_k)} + \frac{1}{(x_j - x_k)(x_j - x_i)}+ \frac{1}{(x_k - x_i)(x_k - x_j)} \right]
\nonumber \\ 
&=\frac{ -(x_j - x_k) - (x_k - x_i) - (x_i - x_j) }{(x_i - x_j)(x_j - x_k)(x_k - x_i)} =0~,
\end{align}
while the last term in Eq.~(\ref{fsq}) can be written as $ \sum_{\substack{i,j = 1\\i\neq j }}^N  {x_i}/{(x_i - x_j)}+ {x_j}/{(x_j - x_i)}= N$. Hence, we finally get
\begin{equation}
\sum_{i=1}^N F_i^2 = \sum_{i = 1}^N \left[ x_i^2 + \sum_{\substack{j = 1\\j\neq i }}^N \frac{1}{(x_i - x_j)^2} \right]-N~, 
\end{equation}
which proves    Eq.~(\ref{vcal_equals_Flog}).

Differentiating Eq.~\ref{vcal_equals_Flog} with respect to $x_j$, we get the force acting on a particle in the Calogero-Moser model:
\begin{equation}
  - \frac{\partial V^{cal}}{\partial x_j} = - \sum_{i=1}^N F_i \frac{\partial F_i}{\partial x_j}~,
\label{force_cal_eqn}
\end{equation}
which vanishes at $x_i=y^{(N)}_i$ since $F_i(y^{(N)}_i)=0$. This proves that the zeros of $H_N(y)$ satisfy Eq.~\ref{min_energy_cal} and thus correspond to the minimum energy configuration of the CM potential.

\subsection{Small oscillations and properties of Hessian matrix}
The Hessian matrix describes small oscillations of a system about its equilibrium configuration ${\bf y}^{(N)}=[y^{(N)}_1,y^{(N)}_2,\ldots,y^{(N)}_N]$, and is defined for the log gas by  
\begin{equation}
    M^{\rm log}_{ij} =  \left[\frac{\partial^2 V^{\rm log}}{\partial x_i \partial x_j} \right]_{{\bf x} = {\bf y}^{(N)}} =\delta_{ij}\left[1+ \sum_{\substack{k=1\\k\neq i}}^N \frac{1}{(y^{(N)}_i - y^{(N)}_k)^2} \right] - (1-\delta_{ij}) \frac{1}{(y^{(N)}_i - y^{(N)}_j)^2}~.
    \label{hess_log_eqn}
\end{equation}
For the CM system, we get,
\begin{equation}
    M^{\rm cal}_{ij} = \delta_{ij}\left[1+ \sum_{\substack{k=1\\k\neq i}}^N \frac{6}{(
 y^{(N)}_i - y^{(N)}_k)^4 } \right] - (1-\delta_{ij}) \frac{6}{(y^{(N)}_i - y^{(N)}_k)^4}~.
    \label{hess_cal_eqn}
\end{equation}
Using the relation Eq.~\ref{vcal_equals_Flog}, it is easy to see that the Hessian matrices of the two models are simply related. Thus, one notes that,
\begin{equation}
   \frac{\partial^2 V^{cal}}{\partial x_j \partial x_k} =  \sum_{i=1}^N \left[\frac{\partial F_i}{\partial x_k} \frac{\partial F_i}{\partial x_j} + F_i \frac{\partial^2 F_i}{\partial x_j \partial x_k} \right] ~.
\end{equation}
Since the forces $F_i$ vanish at the minimum energy configuration ${\bf y}^{(N)}$, we immediately get
$M^{\rm cal}_{j k}= \sum_{i=1}^N {\partial F_i}/{\partial y^{(N)}_k} {\partial F_i}/{\partial y^{(N)}_j} = [(M^{\rm log})^2]_{jk}$ and therefore, we get  the following identity,
\begin{equation}
M^{\rm cal}=(M^{\rm log})^2 \label{HessID}~.
\end{equation}
With this interesting connection between their Hessian matrices, it is expected then that  Gaussian fluctuations and two-point correlation functions  in the two models should be related.  Let us denote, 
\begin{equation}
   S^{(C/L)}_{[i,j]}  = \beta \braket{\delta x_i \delta x_j }~,
   \label{inv hessian element}
\end{equation}
where the superscript $C,L$ indicates CM or LG model respectively.
From the small oscillations theory, we have,
\begin{equation}
 S^{(C/L)}_{[i,j]} = [M^{\rm {cal/log}}]^{-1}~.
\end{equation}
At the level of Gaussian fluctuations we can use the small oscillations approximation and, using the relation in Eq.~(\ref{HessID}), we arrive at the following relation between correlations in the two models: 
\begin{align}
S^C_{i,j} &= \sum_{k=1}^N S^L_{i, k}S^L_{j, k} ~. \nn 
\end{align}
For the LG there are some known exact results on two-point correlations and so 
it may be possible to use the above relation to arrive at some predictions for the CM model. We will explore some of these aspects in the following sections.

Before ending this section we note that the eigenvalues and eigenvectors of the Hessian matrices are known explicitly \cite{calogero1977equilibrium, calogero1981matrices} and we state their explicit forms. Let 
\begin{align}
     H_n(x) &= k_n \prod_{j=1}^{n} (x - x_j) \\
{\rm hence}~~    H'_n(x = x_i) &=\left[ \frac{\partial H_n}{\partial x}\right]_{x = x_i} = k_n \prod_{\substack{j=1\\j\neq i}}^{n} (x_i - x_j)~,
\end{align}
where $H_n(x)$ denotes the Hermite polynomial of order $n$ and $k_n$ is just a normalization constant. A column vector $\mathbf{v}^{(N)}$ is defined such that it's $j^{th}$ element is 
\begin{align}
    \mathbf{v}^{(N)}_j = \frac{1}{H'_N (x_j)}~,~~j = 1,2,\ldots,N~,
    \label{basis_vec}
\end{align}
while $\mathbf{X}^{(N)}$ is defined as a diagonal matrix with diagonal elements being set to be the $N$ zeros of $H_N(x)$.  Then the eigenvectors of $M^{\rm log}$ (and naturally of $M^{\rm cal}$) are given by  
\begin{equation}
    \bm{\psi}^{(N)}_j = H_{j-1}(\mathbf{X}^{(N)}). \mathbf{v}^{(N)},~~ j = 1,2,\ldots, N~.
    \label{eigvectors_of_hess}
\end{equation}
The eigenvalue corresponding to the $j^{\rm th}$ eigenvector   $\bm{\psi}^{(N)}_j$ is $j$ for  $M^{\rm log}$ and $j^2$ for  $M^{\rm cal}$.



\section{Numerical  results for equilibrium properties}
\label{eqb}

We now present various results on equilibrium properties of the   two systems obtained from  direct Monte-Carlo simulations.  Most of our results are for the values $\beta=1,2,4$ of the inverse  temperature.

We are particularly interested in: 

(i) The particle density profile 
\begin{equation}
P(x)=\frac{1}{N} \sum_{i=1}^N \langle \delta (x-x_i) \rangle~,
\end{equation}
where $\langle ... \rangle$ denotes an equilibrium average over the distribution $e^{-\beta V({\bf x})}/Z$. 

(ii) The marginal distributions
\begin{equation}
P_k(x)= \langle \delta (x-x_k) \rangle~,
\end{equation}
where $k=N$ (or $k=1$) refers to the edge particle and $1<<k<<N$ refers to bulk particles and the associated moments and cumulants of these distributions (mean, variance, skewness and kurtosis).  

(iii) The two-point correlation functions $ \langle \delta x_i \delta x_j \rangle$.

We also compute various properties from the small oscillations approximation. The small oscillation theory would predict that the mean position of the $k$-th particles is simply given by $y_k$ while correlation functions are given by the inverse of the Hessian matrix
\begin{equation}
M^{-1}_{i,j} :\equiv S_{[i,j]}~. 
\end{equation}
Note that this gives mean-squared-deviation of the $k$-th particle to be $S_{[k,k]}$. 

{\emph{Simulation approach}}: Note that here we are mostly interested in the properties of the system where particles maintain their ordering. This would normally require us to perform Monte-Carlo moves (single-particle displacements) of very small sizes so that particle-crossings are avoided, and this would make equilibration very slow.  We avoid this issue by ignoring particle crossings during most of the Monte-Carlo steps but then ordering the particles during the data-collecting steps. The step-lengths  were chosen so that acceptance was around $50 \%$.   In our simulatons we collect data after every $5$ Monte-Carlo cycles (each cycle involving $N$ attempts) and averages were typically computed with  around $10^7-10^8$ samples. For marginal distributions and cumulants, the averages were computed over a larger ensemble with around $10^{10}$ samples for greater precision. Some checks on equilibration are shown in the appendix.

\subsection{Density profile}
\begin{figure}[!h]
    \begin{subfigure}[b]{0.49\textwidth}
        \includegraphics[width=\linewidth]{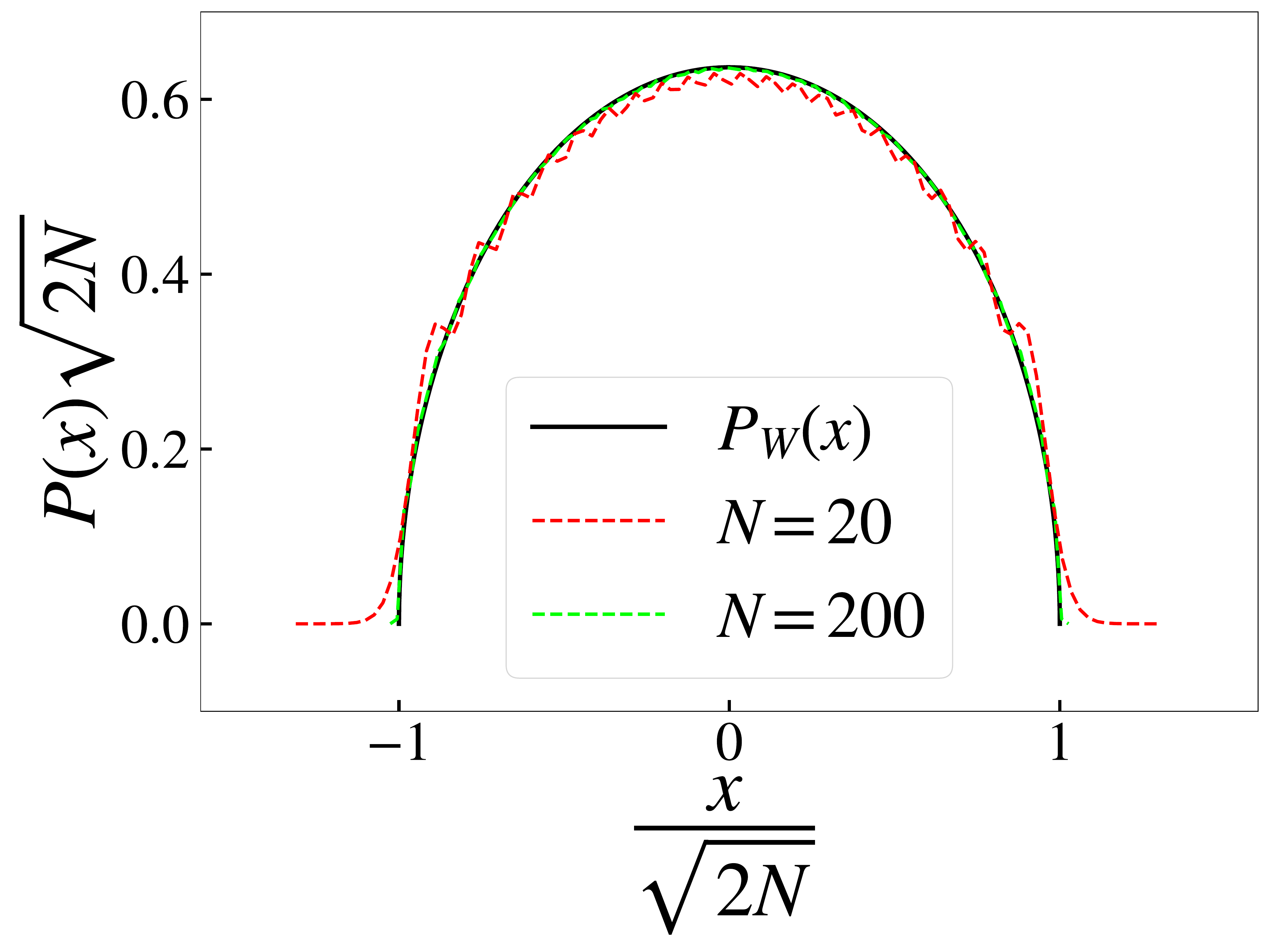}
        \caption{}
        \label{fig:Wigner semi circle cal}
    \end{subfigure}
    \hfill 
    \begin{subfigure}[b]{0.49\textwidth}
        \includegraphics[width=\linewidth]{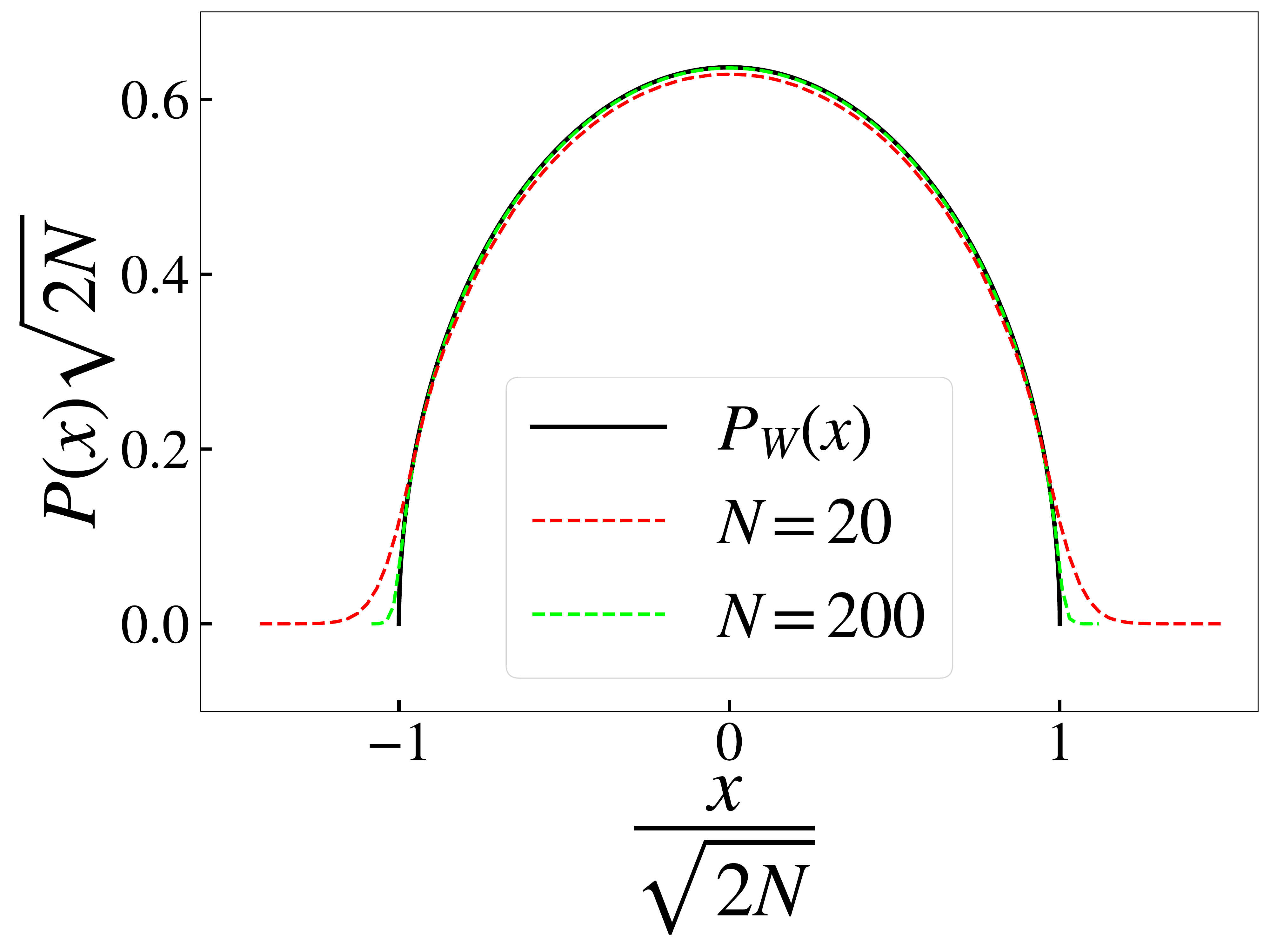}
        \caption{}
        \label{fig:Wigner semi circle log}
    \end{subfigure}
    \caption{$P(x)$ versus $x$: Density of particles for $\beta$ = 4 for: (a) Calogero Model (b) Log-gas }
\end{figure}
Since the LG and CM have minimum energy configurations given by the Hermite polynomial zeros, it is expected that for low temperatures and in the thermodynamic limit, both the models will have single-particle distributions described by the Wigner semi-circle \cite{wigner1951statistical}
\begin{equation}
P_W(x) = \frac{1}{\pi}\sqrt{2N - x^2}~. 
\end{equation} 
For the LG this is well known from random matrix theory and 
it has also been shown to be true for  the CM model \cite{abanov2011soliton,nadal2011TWderiv}.  We show in Figs.~(\ref{fig:Wigner semi circle cal},\ref{fig:Wigner semi circle log}), results from our simulations for $P(x)$ where we find that the agreement with $P_W(x)$ is  good for finite systems even at high temperatures. We observe some deviations from the semicircle at the edges and this is due to finite-size effects. Note also the oscillations around the mean density profile that can be seen for the CM case. These oscillations, which decrease with increasing system size, indicate that the CM system is more rigid and that oscillations of particles around their mean position is smaller than in the LG.

\subsection{Single-particle marginal distributions}

\subsubsection{Statistics of the edge particle position $x_N$} 
\begin{figure}[!ht]
    \begin{subfigure}[b]{0.49\textwidth}
        \includegraphics[width=\textwidth]{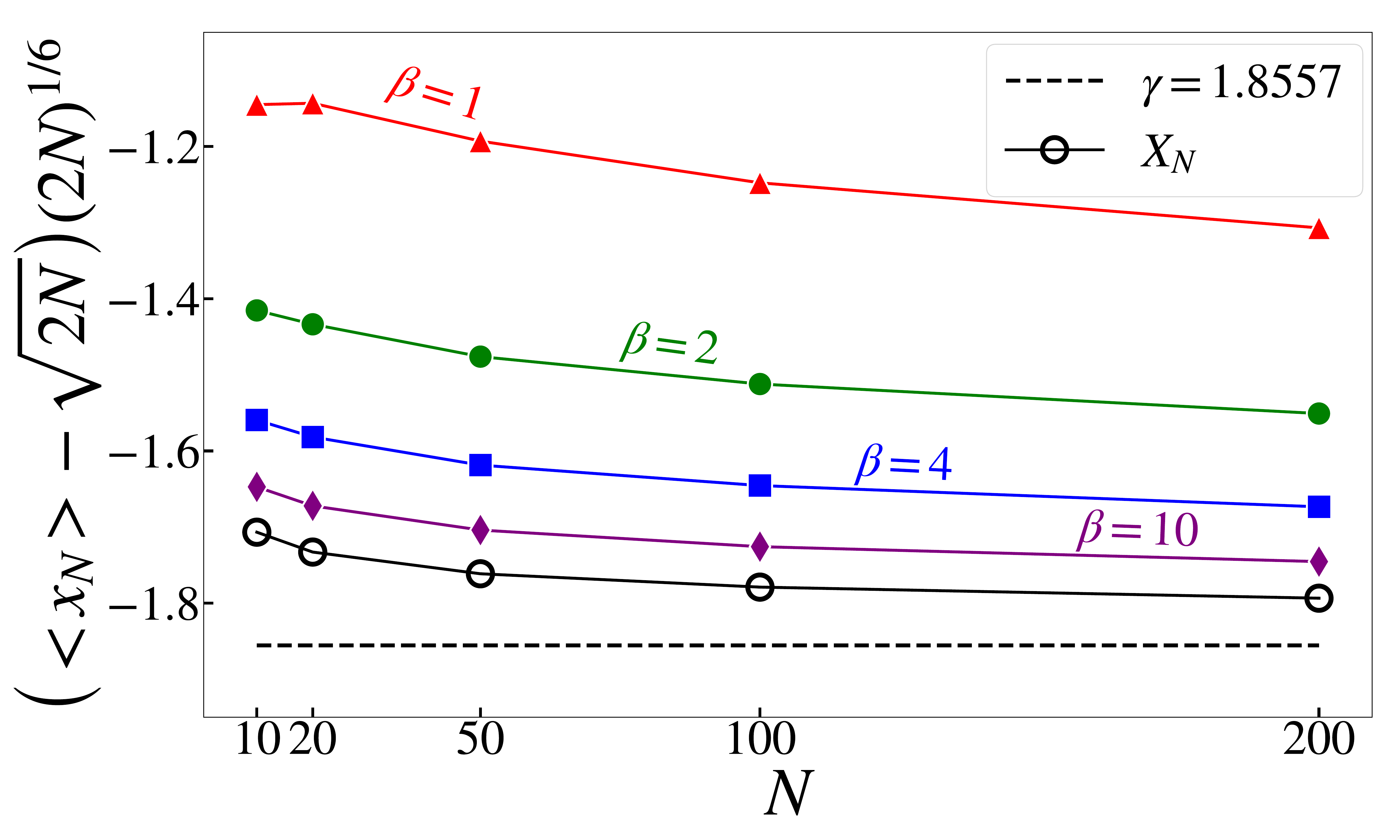}
        \caption{}
        \label{fig:mean of x_N cal}
    \end{subfigure}
    \hfill 
    \begin{subfigure}[b]{0.49\textwidth}
        \includegraphics[width=\textwidth]{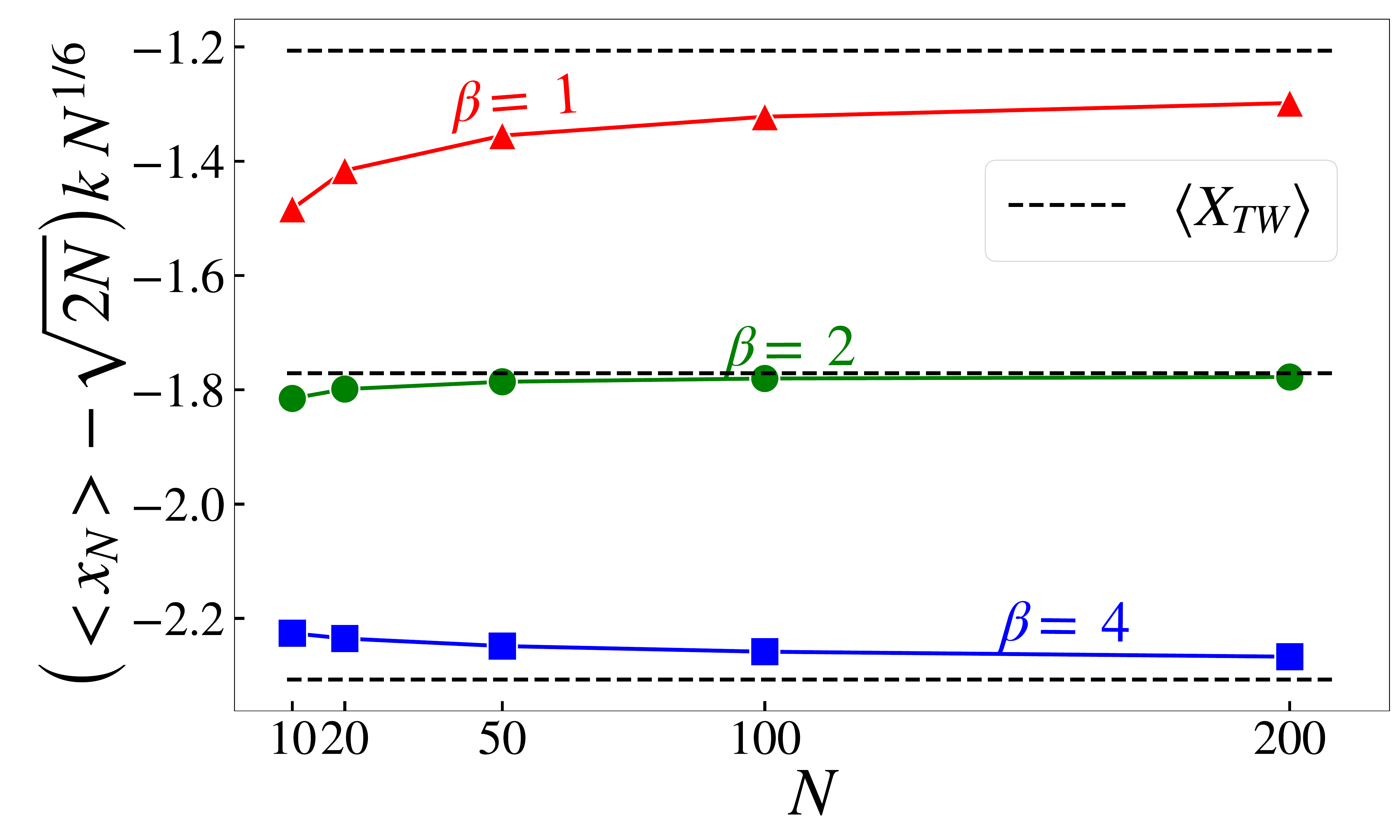}
        \caption{}
        \label{fig:mean of x_N log}
    \end{subfigure}
    \newline 
    \caption{ Plot of scaled mean displacement, from $\sqrt{2 N}$,  of the edge particle $x_N$, as a function of $N$,  for (a) the CM model and (b) the LG model. For (a) we see that the low-temperature asymptotic value is close to the dashed line which is the value expected for the $N-$th zero of $H_N(y)$ in the large-$N$ limit. The rings, which denote exact values of the $N-$th zero of $H_N(y)$, approach the dashed line at very large-$N$. For (b) the asymptotic values are seen to approach the values expected from the TW distributions for $\beta=1,2,4$.}   
\end{figure}

\begin{figure}[!ht]
    \begin{subfigure}[b]{0.49\textwidth}
        \includegraphics[width=\textwidth]{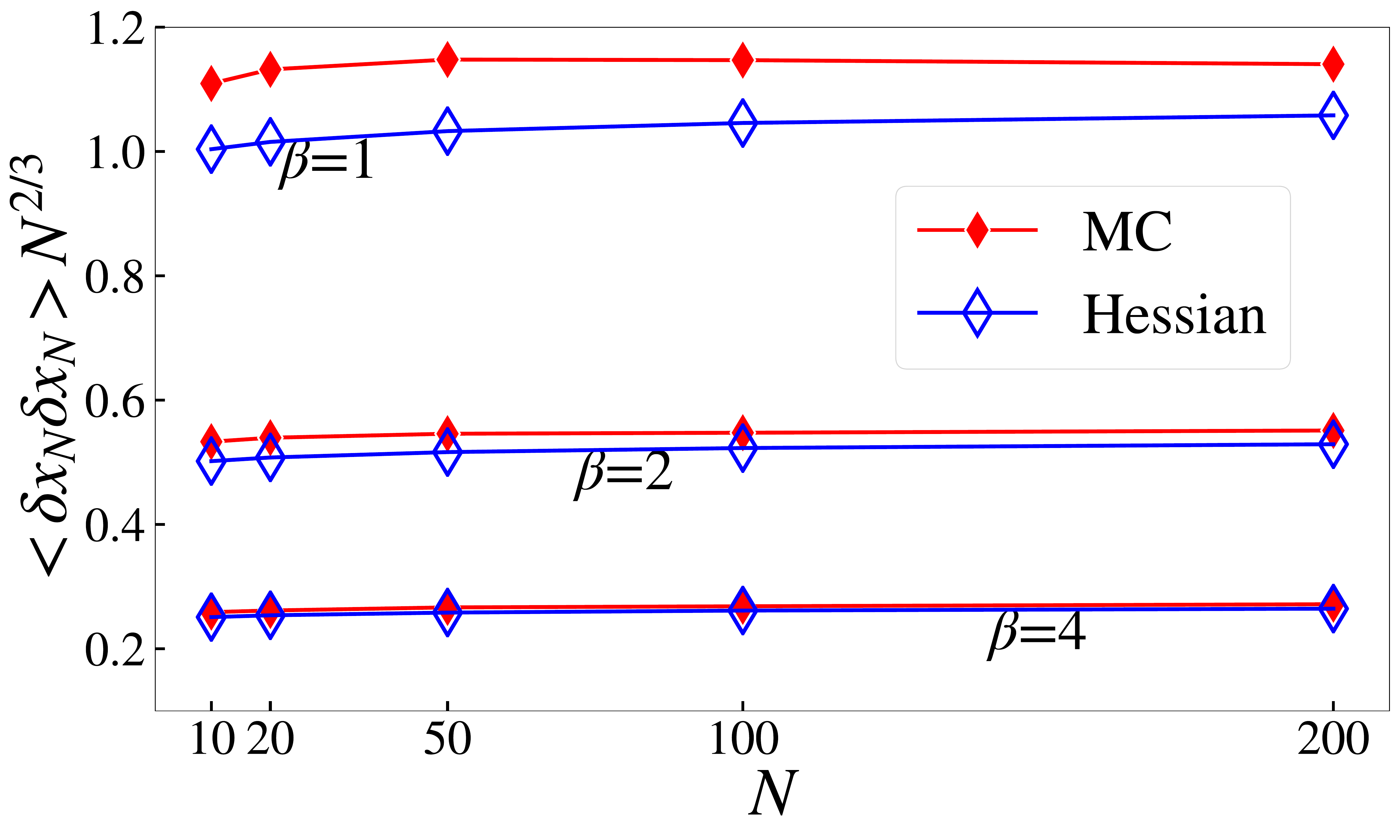}{\tiny {\normalsize {\normalsize }}}
        \caption{}
       \label{fig:MSD of x_N cal}
    \end{subfigure}
    \hspace*{\fill} 
    \begin{subfigure}[b]{0.49\textwidth}
        \includegraphics[width=\textwidth]{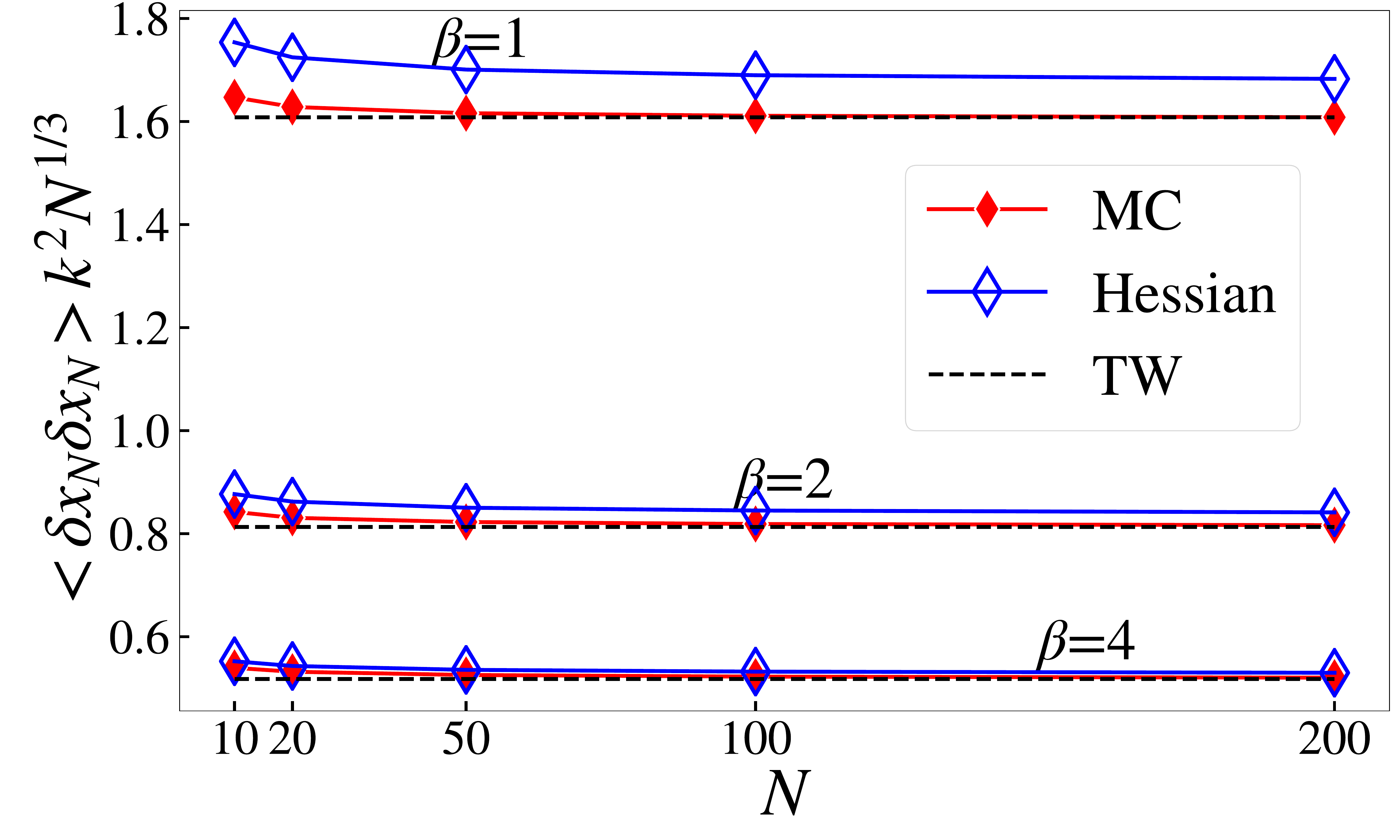}
        \caption{}
        \label{fig:MSD of x_N log}
    \end{subfigure}
\caption{Plot of the scaled MSD of the edge particle $x_N$ plotted as a function of $N$ for (a) the CM model and (b) the LG model. The results from direct simulations and from the Hessian approach are shown and we see that they start differing at higher temperatures. However both predict the scaling $\langle \delta x_N^2 \rangle \sim N^{-2/3}$ in the CM model and larger fluctuations $\langle \delta x_N^2 \rangle \sim N^{-1/3}$ in the LG case. For the LG case we find the expected convergence of the MSD to the values obtained from  TW.}
\end{figure}

--- In Figs.~(\ref{fig:mean of x_N cal},\ref{fig:mean of x_N log},\ref{fig:MSD of x_N cal},\ref{fig:MSD of x_N log}) we show results for the mean position $\langle x_N \rangle$ and the variance $\langle \delta x_N^2 \rangle$ of the edge particle. These are compared with the small-oscillation approximation results $y_N$ and $S(N,N)$ for the mean and variance respectively. From the properties of the Hermite zeros, it is known that 
the asymptotic value of $y_N$ is given by~\cite{szeg1939orthogonal}:
\begin{equation}
    y_N = \sqrt{2N} - \frac{\gamma}{(2N)^{1/6}}
    \label{nth eigenvalue}
\end{equation}
where, $\gamma = 1.8557$. From Figs.~(\ref{fig:mean of x_N cal},\ref{fig:mean of x_N log}) we see that for both the CM and LG model, the mean has the form 
\begin{equation}
\langle x_N \rangle = \sqrt{2N}-\frac{c}{N^{1/6}}~, 
\end{equation}
where the constant $c$ is temperature dependent, and approaches the constant $\gamma$, in the $\beta \to \infty$ limit. For the log-gas, it is known that the scaled variable 
\begin{equation}
z=(x_N  - \sqrt{2N})~k N^{1/6}~,\label{scalxN}
\end{equation}
with $k=2^{1/2}$ for $\beta=1,2$ and $k=2^{2/3}$ for $\beta=4$, satisfies the Tracy-Widom distribution $TW_\beta(z)$. Hence for the edge-particle mean in the log-gas, the limiting result  $\lim_{N \to \infty} \langle z \rangle \to \langle z \rangle_{TW}$ is expected. Our numerics in Fig.~(\ref{fig:mean of x_N log}) verify this.

\begin{figure}[!h]
	\begin{subfigure}[b]{0.47\textwidth}
		\includegraphics[width=\linewidth]{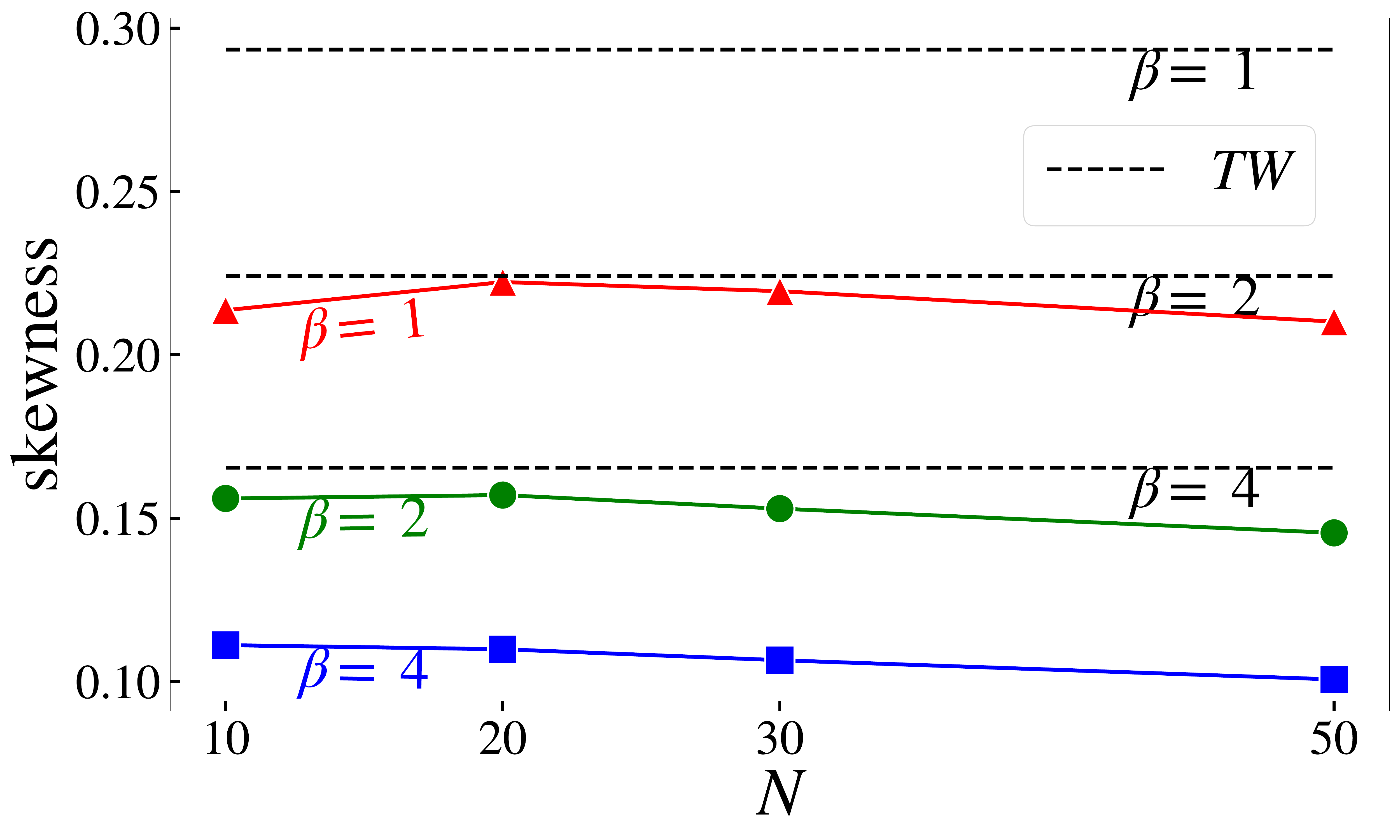}
		\caption{}
		\label{fig:skew of x_N CM}
	\end{subfigure}
	\hfill
	\begin{subfigure}[b]{0.47\textwidth}
		\includegraphics[width=\linewidth]{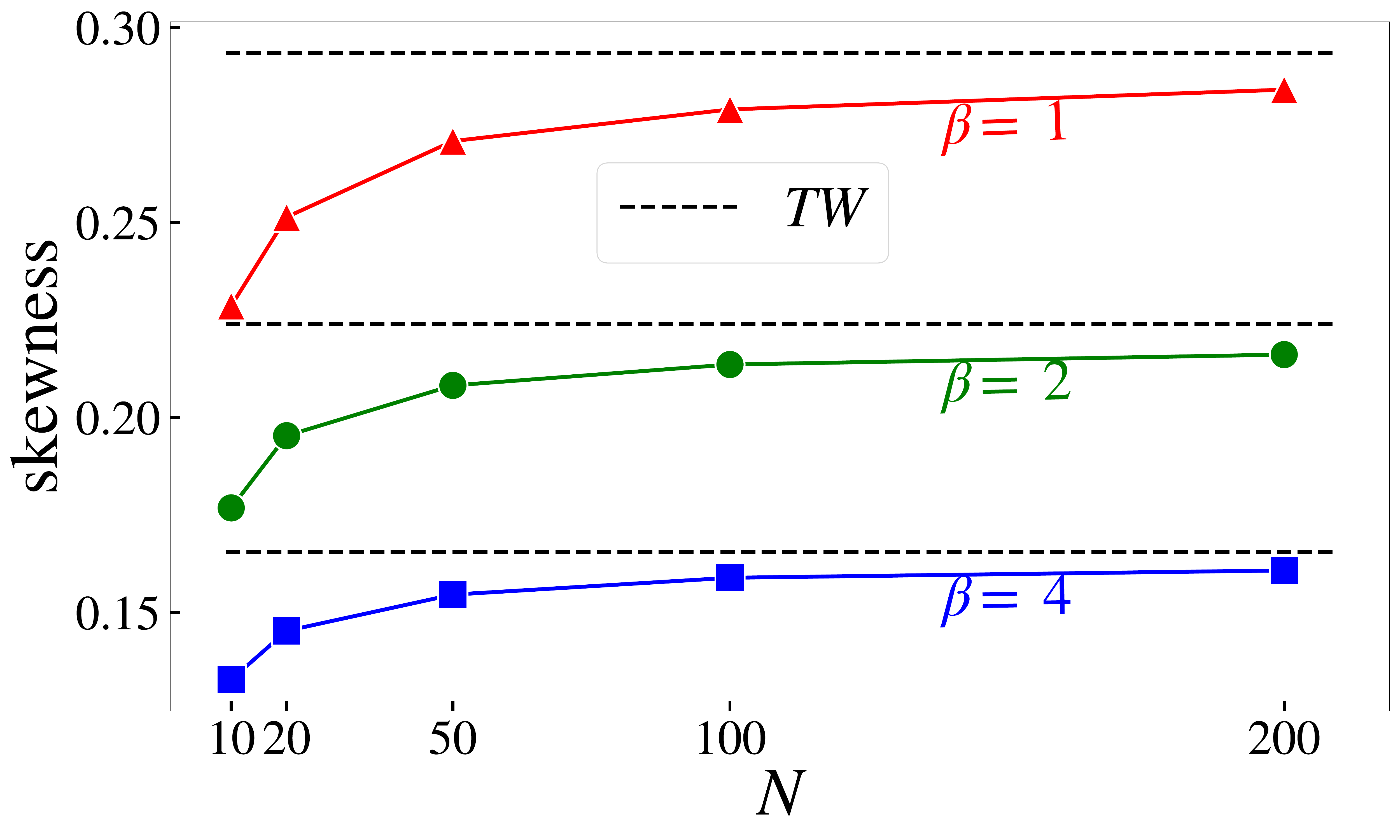}
		\caption{}
		\label{fig:skew of x_N log}
	\end{subfigure}
	\newline
	\begin{subfigure}[b]{0.47\textwidth}
		\includegraphics[width=\linewidth]{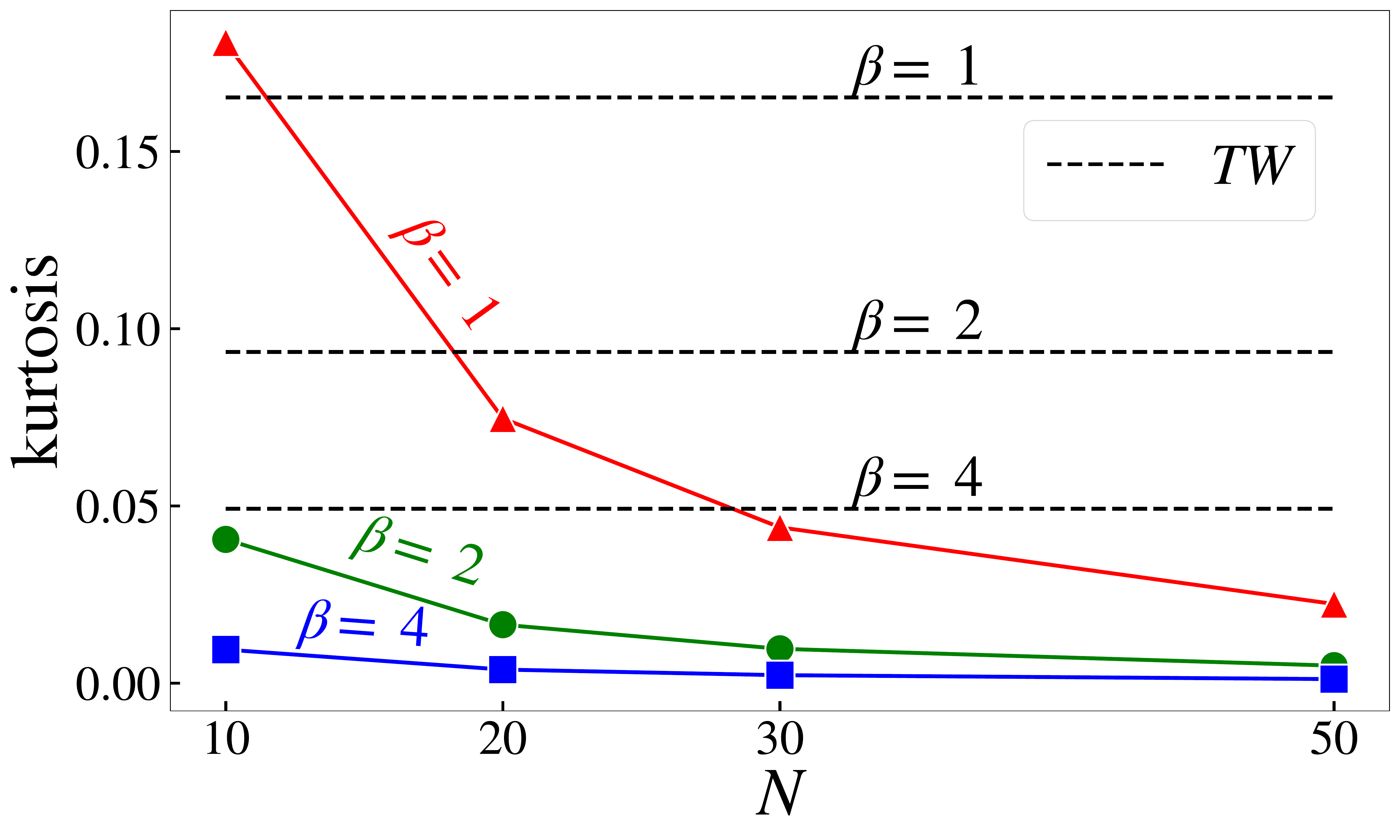}
		\caption{}
		\label{fig:kurt of x_N CM}
	\end{subfigure}
	\hfill
	\begin{subfigure}[b]{0.47\textwidth}
		\includegraphics[width=\linewidth]{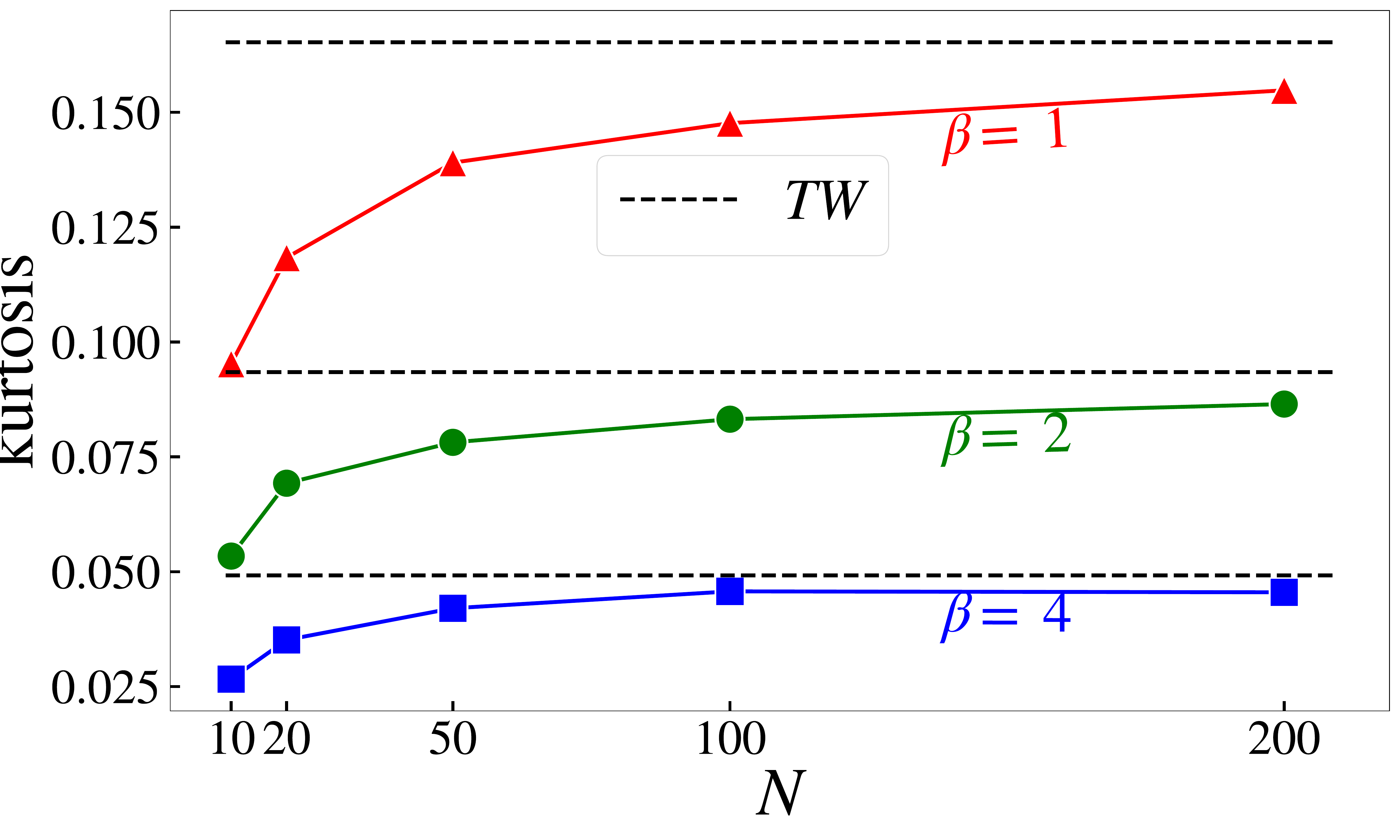}
		\caption{}
		\label{fig:kurt of x_N log}
	\end{subfigure}
		\caption{Figures (a) and (b) show plots of skewness versus system size for the CM  and LG models respectively while figures (c) and (d) show plots of the kurtosis for the two cases. For the LG, we get the expected asymptotic values predicted by TW, while for the CM case it is clear that they differ significantly. }
\end{figure}

\begin{figure}[!ht]
    \begin{subfigure}[b]{0.49\textwidth}
        \includegraphics[width=\linewidth]{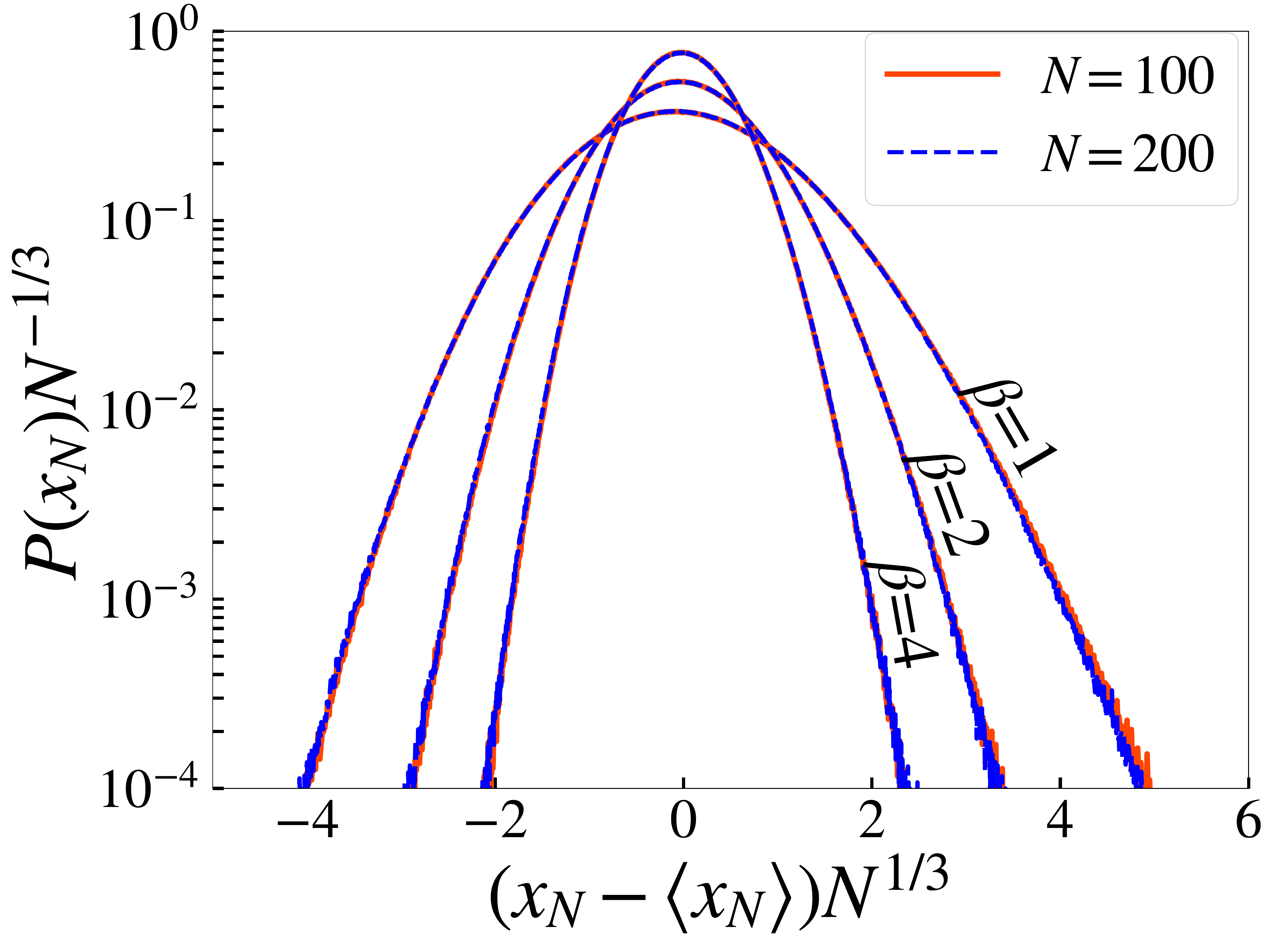}
        \caption{}
        \label{fig:TW with x_N cal}
    \end{subfigure}
    \hfill 
    \begin{subfigure}[b]{0.49\textwidth}
        \includegraphics[width=\linewidth]{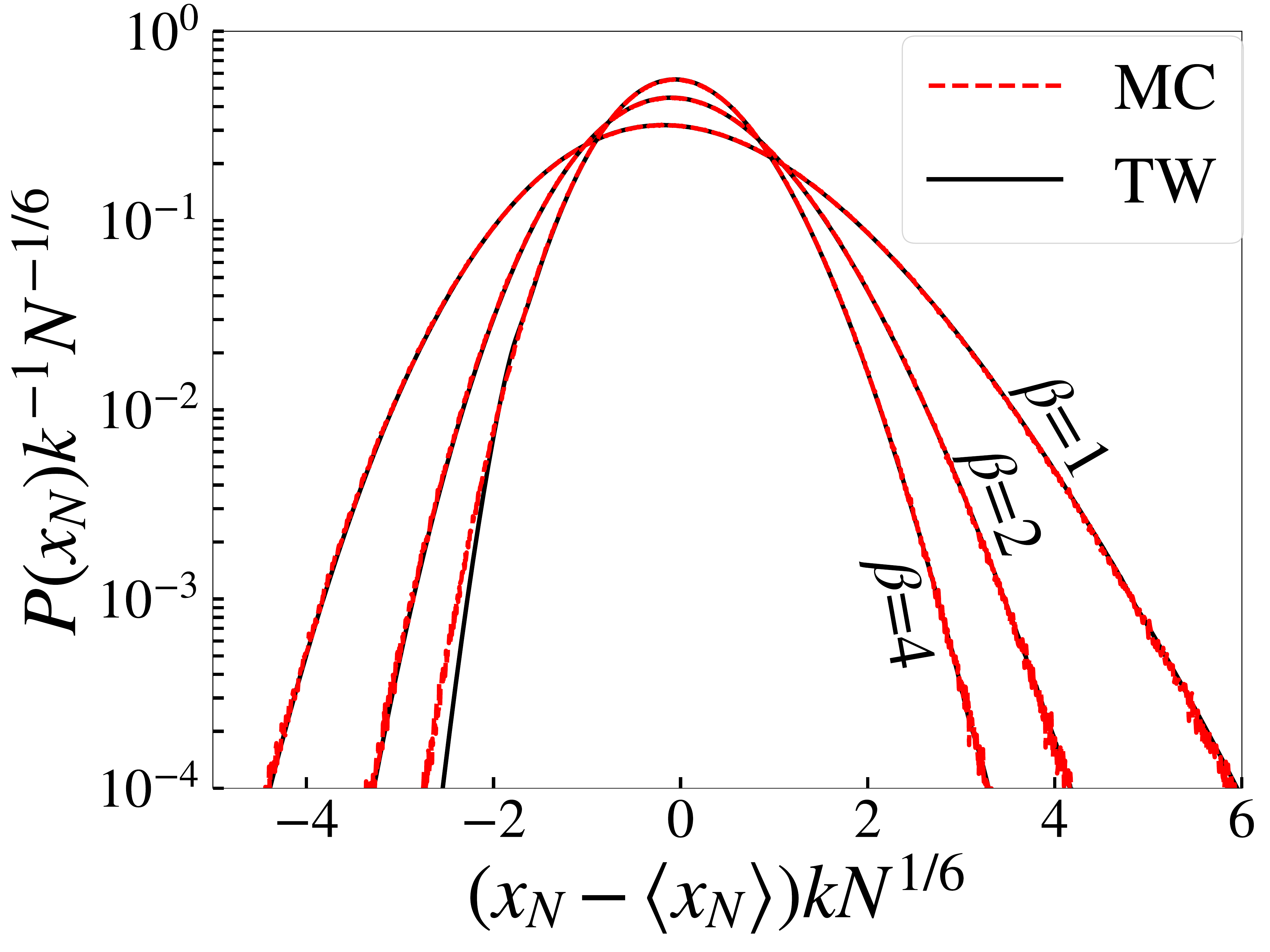}
        \caption{}
        \label{fig:TW with x_N log}
    \end{subfigure}
    \caption{Full probability distribution of edge particle in (a) the CM model and (b) the LG, with appropriate scale factors. In (a) we show results for two system sizes $N=100,200$, and see  good convergence indicating that we are already in the scaling limit. For (b) we only show the $N=200$ data since the $N=100$ scaled data is indistinguishable. We also find excellent agreement with the Tracy-Widom distribution over a wide range. The scale factors $k = 2^{1/2}$ for $\beta = 1, 2$ and $k = 2^{2/3}$ for $\beta = 4$ are the standard ones for TW.} 
\end{figure}

The fluctuations are much smaller in the CM model as compared with the LG. As shown in Figs.~(\ref{fig:MSD of x_N cal},\ref{fig:MSD of x_N log}),  we find that 
\begin{equation}
\langle \delta x_N^2 \rangle \sim N^{-2/3}
\end{equation}
for the CM, while  the LG gives 
\begin{equation}
\langle \delta x_N^2 \rangle \sim N^{-1/3}~.
\end{equation}
We have also plotted the results from the Gaussian theory and we see that these also reproduce the correct scalings, though not the precise prefactors. For the LG we have $ \langle (z- \langle z \rangle)^2 \rangle =  (x_N  - \langle x_N \rangle)^2~k^2 N^{1/3}$ and this is again known from the variance of the Tracy-Widom. We verify this in Fig.~(\ref{fig:MSD of x_N log}). A natural question is whether one can define an appropriately shifted and  scaled variable, as in Eq.~(\ref{scalxN}), which would satisfy the Tracy-Widom distribution. Since we do not have a theory which tells us what the shift and scale factors should be, it is not possible to test this from results on the mean and variance. However, the skewness and kurtosis of the distribution are 
quantities which are independent of both the shift and scale factors and computing these for the CM model gives us a direct way to test possible connections to Tracy-Widom.  In Figs.~(\ref{fig:skew of x_N CM},\ref{fig:skew of x_N log}), we plot results for the skewness in the CM and LG respectively while in  Figs.~(\ref{fig:kurt of x_N CM},\ref{fig:kurt of x_N log})  we plot the kurtosis in the two models. We find that while the LG results are consistent with that expected from Tracy-Widom, in the CM model we find that both quantities seem to decay with increasing system size, implying that the fluctuations are approximately Gaussian.

 Full distribution  of the edge particle $P_N(x)$: In Figs.~(\ref{fig:TW with x_N cal},\ref{fig:TW with x_N log}), we plot the full distributions of the edge particle position for the two models. For the LG model we find a very accurate verification of the TW distribution. On the other hand, for the CM model, the typical fluctuations appear to be Gaussian while the large deviations show significant asymmetry.

\subsubsection{Statistics of bulk particle position $x_{N/2}$}

For the log-gas, two-point correlations of the ordered particles have been computed exactly \cite{o2010gaussian} and it has been shown that bulk correlations show  Gaussian fluctuations. We summarize here the main results. It was shown that
the the mean squared deviation of a bulk particle $1<<k<<N$ is given by
\begin{equation}
    \braket{\delta x_k \delta x_k} = \frac{\log N}{2\beta [1-t^2(k)]N}
    \label{MSD_theor}
\end{equation}
where $t(k)$ is to be found by inverting the relation
\begin{equation}
    \frac{k}{N} = \frac{2}{\pi}\int_{-1}^t \sqrt{1-x^2}  dx~.
\end{equation}
It was also shown that correlations of bulk particles separated by distance 
$O(N)$ decay faster than $\log (N)/N$. From our numerics of the LG, we verify that the MSD of bulk particles scale as $(\log N)/N$. We also find that long-range correlations  decay as $1/N$ (see next section). On the other hand for the CM, both the MSD and correlations decay as $1/N$. We now present the numerical results. 
\begin{figure}[!ht]
    \begin{subfigure}[b]{0.52\textwidth}
        \includegraphics[width=\linewidth]{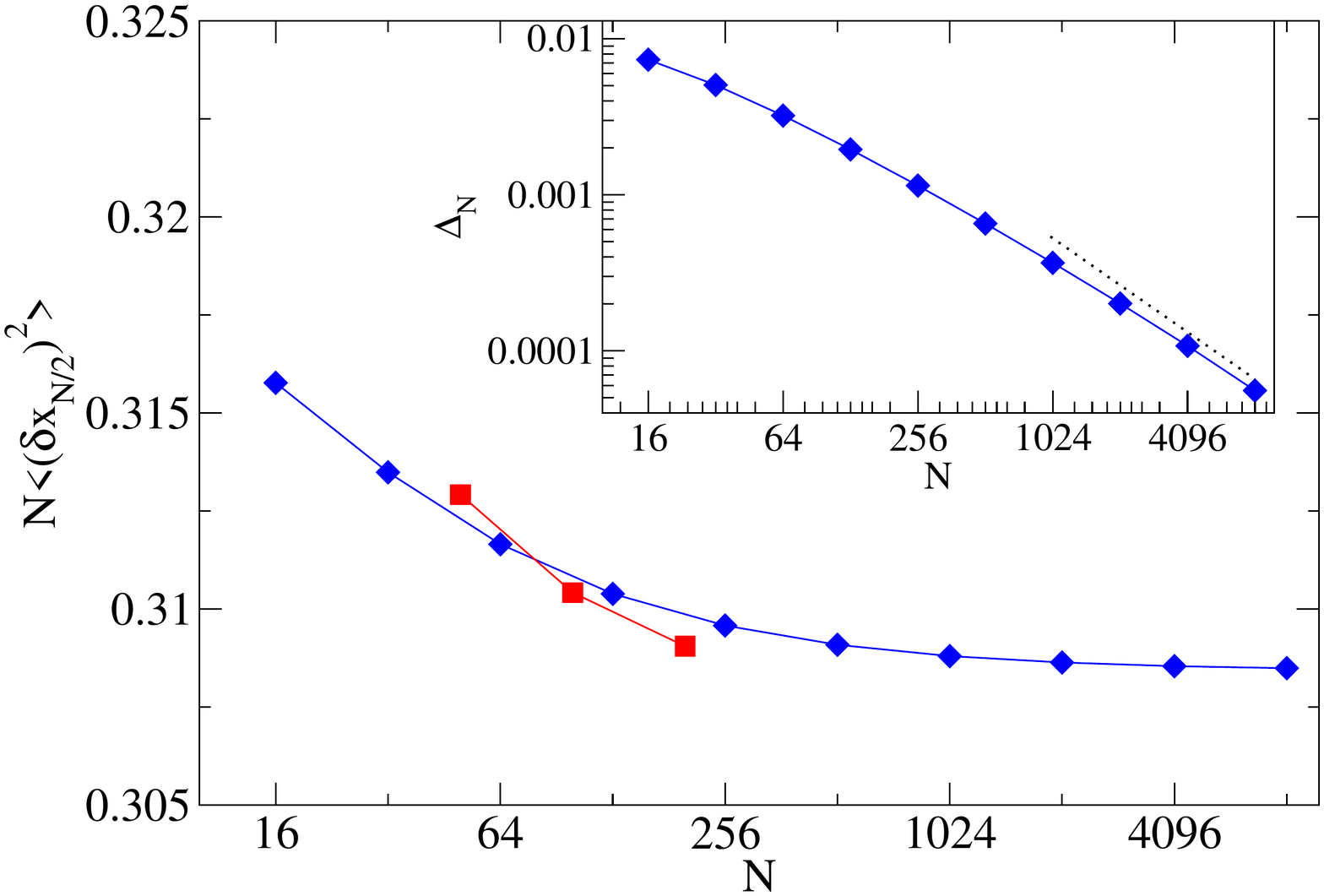}
        \caption{}
        \label{MSD-CM-Nby2}
    \end{subfigure}
    \hfill
    \begin{subfigure}[b]{0.52\textwidth}
       \includegraphics[width=\linewidth]{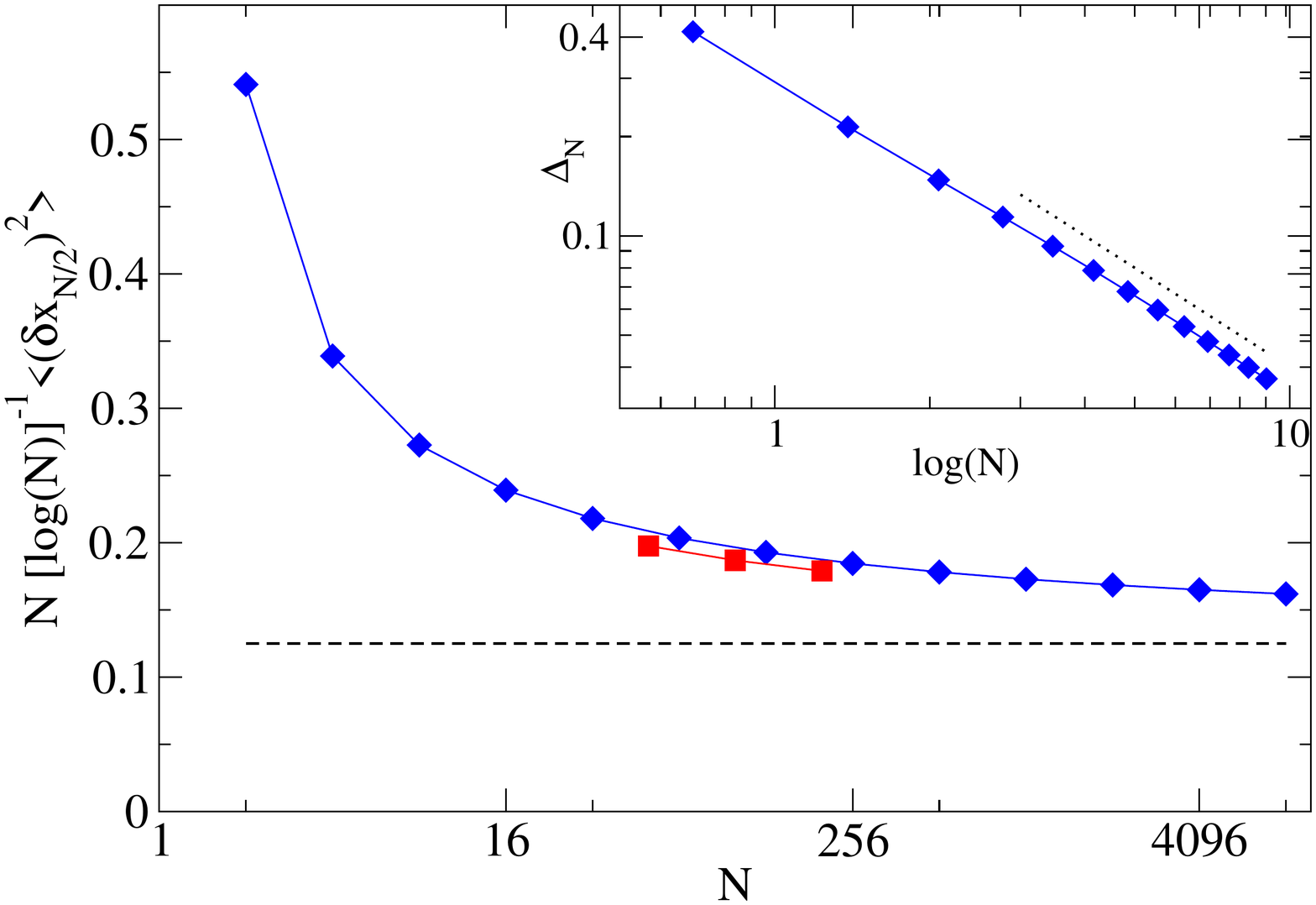}
        \caption{}
        \label{MSD-LG-Nby2}
    \end{subfigure}
    \hfill
    \caption{The scaled mean-squared-deviation in the position of  particle  $k=N/2$, as obtained from simulations (red squares) and  from the Hessian (blue diamonds) for (a) the CM model and (b) the LG model, for $\beta=4$.     
For the LG model, we also show the expected theoretical prediction from Eq.~(\ref{MSD_theor}) with the dashed line. For the CM model,  the data fits well  to the form $A+B/N$ with $A=0.3084$ and in the inset we plot, in log-log scale, $\Delta_N=N \langle \delta x_{N/2}^2 \rangle- A$, which shows the $N^{-1}$ decay (dotted line).   
The inset in  (b) in log-log scale shows the difference $\Delta_N=[N/\ln (N)]  \langle \delta x_{N/2}^2 \rangle -1/8$ (theoretical value) and we see a $[\log (N)]^{-1}$ decay (dotted line).}  
\end{figure}


\begin{figure}
	\begin{subfigure}[b]{0.49\textwidth}
		\includegraphics[width=\linewidth]{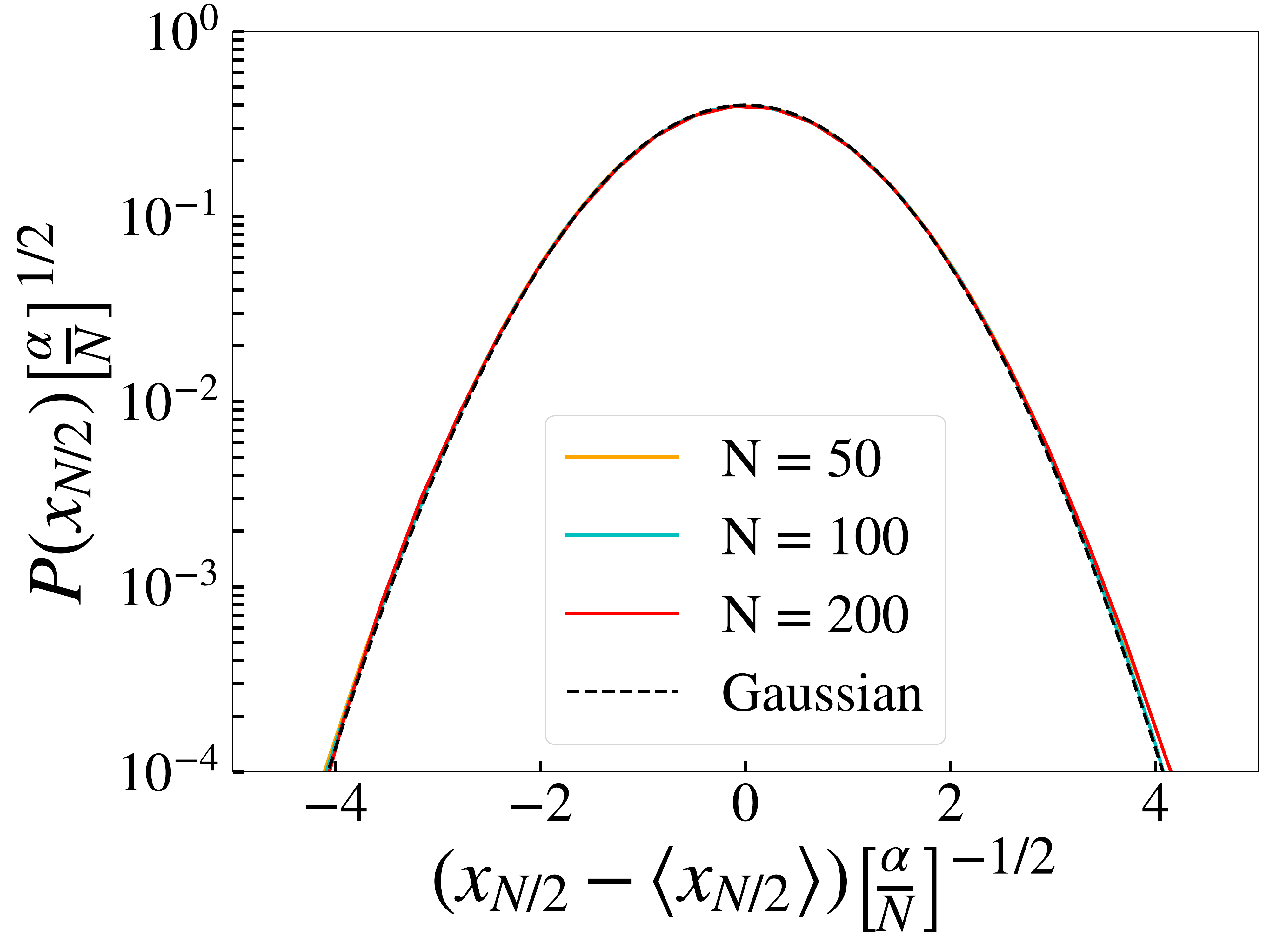}
		\caption{}
        \label{fig:prob_xNby2_CM}
	\end{subfigure}
	\hfill
    \begin{subfigure}[b]{0.49\textwidth}
        \includegraphics[width=\linewidth]{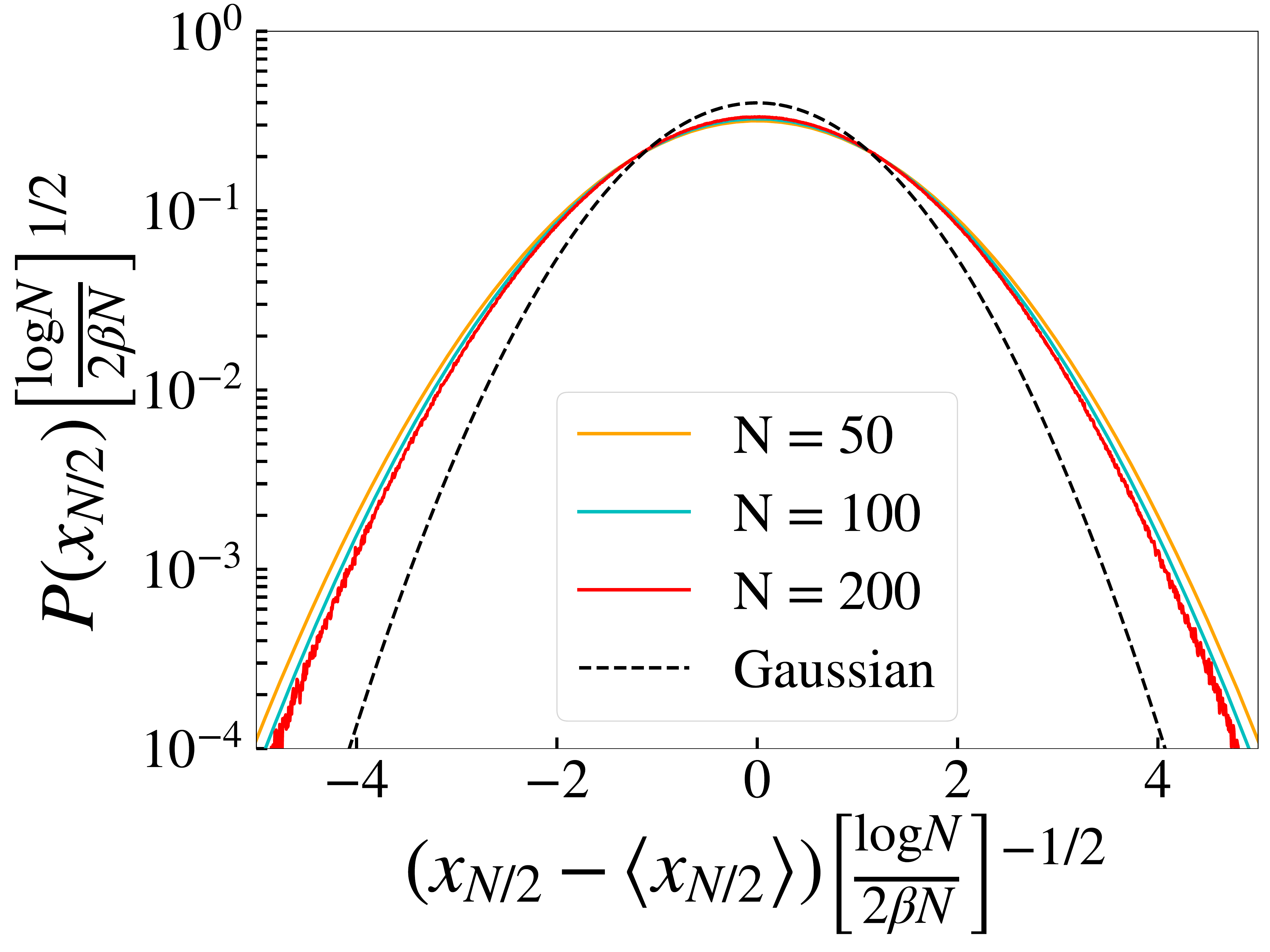}
        \caption{}
        \label{fig:prob_xNby2_LG}
    \end{subfigure}
    \hfill
    \caption{Probability distribution of $x_{N/2}$ obtained from simulations, for (a) CM model and (b) LG model, for three different system sizes  $N=50,100,200$.   For (a) we plot a Gaussian corresponding to the asymptotic MSD  [see Fig.~(\ref{MSD-CM-Nby2})] where $\alpha$ = A = 0.3084, and  see a fast convergence of the simulation results to this form. For (b), we see a slow convergence to the expected theoretical Gaussian form for the LG.}
\end{figure}

 In Figs.~(\ref{MSD-CM-Nby2},\ref{MSD-LG-Nby2}), we plot the size-dependence of the mean-squared-deviations in the position of the $N/2$-th  particles, for the CM and LG models respectively for $\beta =4$.  Results of both direct simuations for sizes $N=50,100,200$, and the Hessian theory (where much larger sizes can be studied), are shown and we see good agreement between the two. 
 For the CM model, there are no theoretical predictions and we find in Fig.~(\ref{MSD-CM-Nby2}) that $\langle \delta x_{N/2}^2 \rangle$ scales as $1/N$.   In fact our data fits well to the form  $N \langle \delta x_{N/2}^2 \rangle = A+B/N$ with $A\approx 0.3084$ and the the inset shows a $\sim 1/N$ convergence to this asymptotic value.  For the LG gas,  using Eq.~(\ref{MSD_theor}) with  $k=N/2$ and $t(k=N/2)=0$, gives $N\langle \delta x_{N/2}^2 \rangle/\log(N) = 1/8$. We see in the inset of Fig.~(\ref{MSD-LG-Nby2}) a slow convergence ($\sim1/\log N$) to this limiting value. In Figs.~(\ref{fig:prob_xNby2_CM},\ref{fig:prob_xNby2_LG}) we plot the full probability distributions, appropriately scaled, of the $N/2$-th particle. For the CM model we show the Gaussian form using the asymptotic variance $N \langle \delta x_{N/2}^2 \rangle=0.3084$. For the LG we again see a slow convergence to the theoretically expected Gaussian distribution. 
 
\begin{figure}[!h]
    \begin{subfigure}[b]{0.49\textwidth}
        \includegraphics[width=\linewidth]{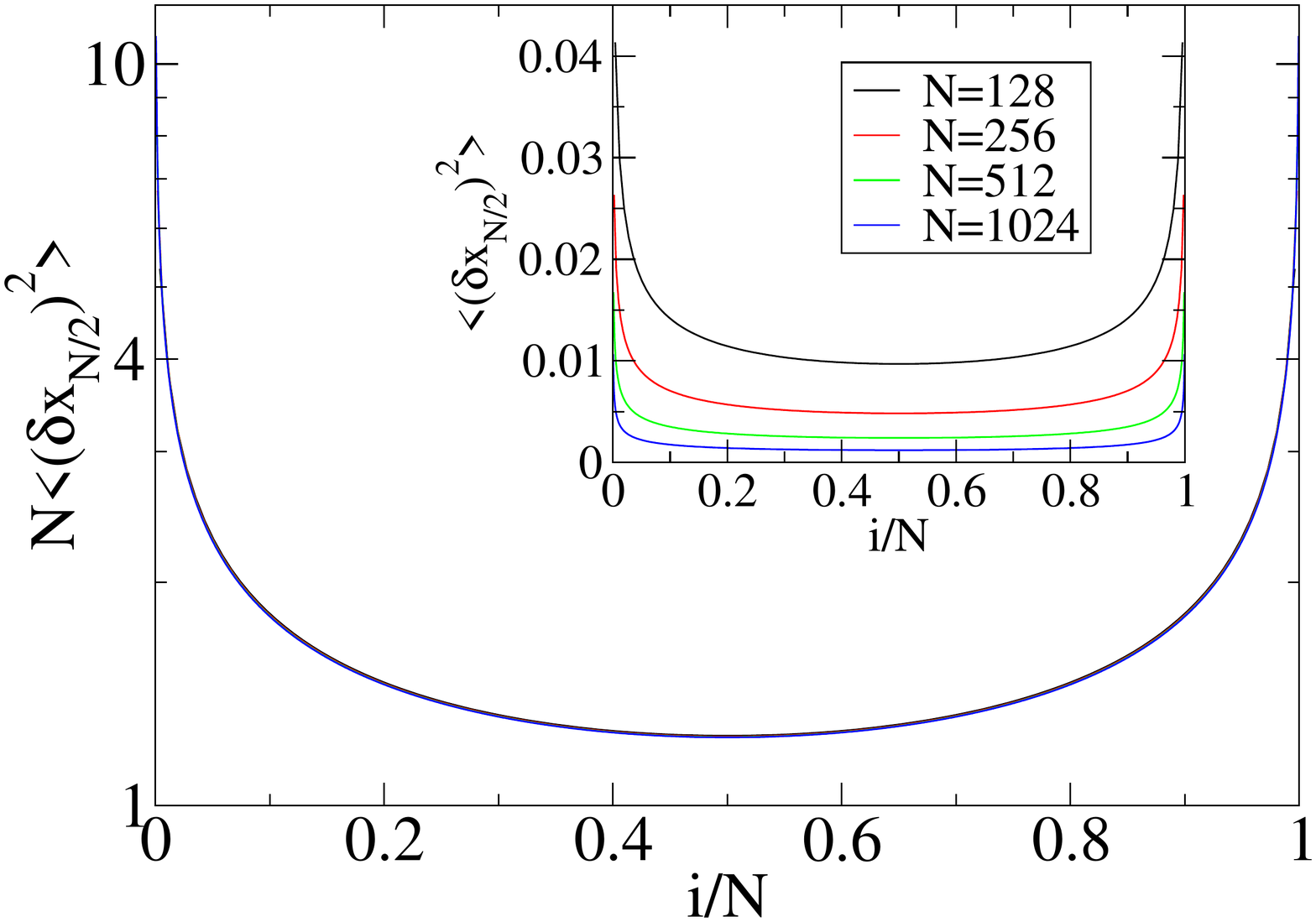}
        \caption{}
\label{CMSii}
    \end{subfigure}
    \begin{subfigure}[b]{0.49\textwidth}
        \includegraphics[width=\linewidth]{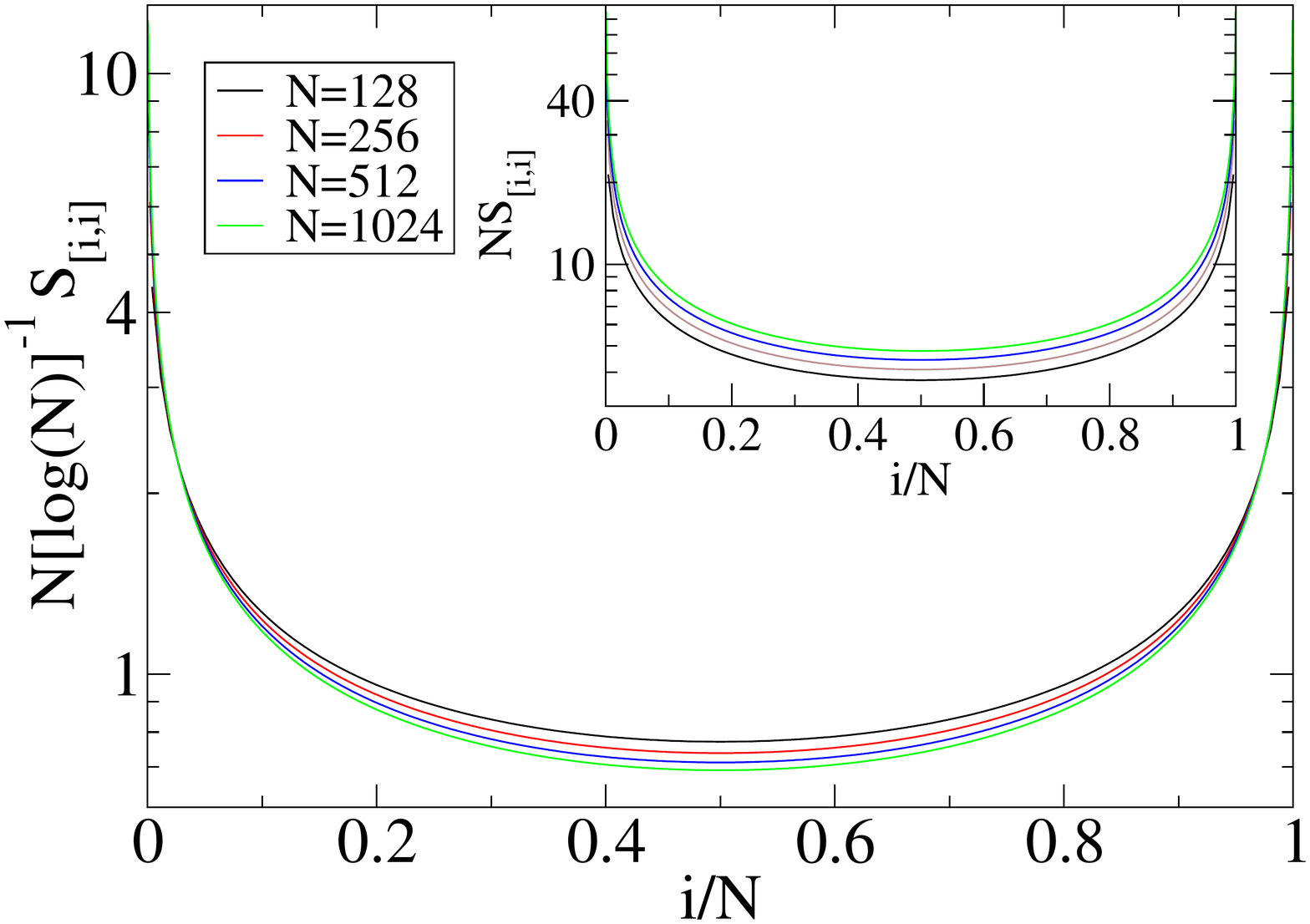}
        \caption{}
\label{LGSii}
    \end{subfigure}
\caption{ MSD profiles in (a) the CM and (b) the LG model, as computed from the Hessian, for different system sizes. In (a) we see a very good $1/N$ scaling while in (b) we show that a $\ln(N)/N$ scaling gives a better collapse than a $1/N$ scaling (inset).}
\end{figure}

In Figs.~(\ref{CMSii},\ref{LGSii}), we show that the mean squared displacements at all bulk points, evaluated from the Hessian, also satisfy the $1/N$ and $\log(N)/N$ scaling for the CM and LG systems respectively.

\subsection{Results on two-point correlations}

\begin{figure}[!h]
    \begin{subfigure}[b]{0.49\textwidth}
        \includegraphics[width=\linewidth]{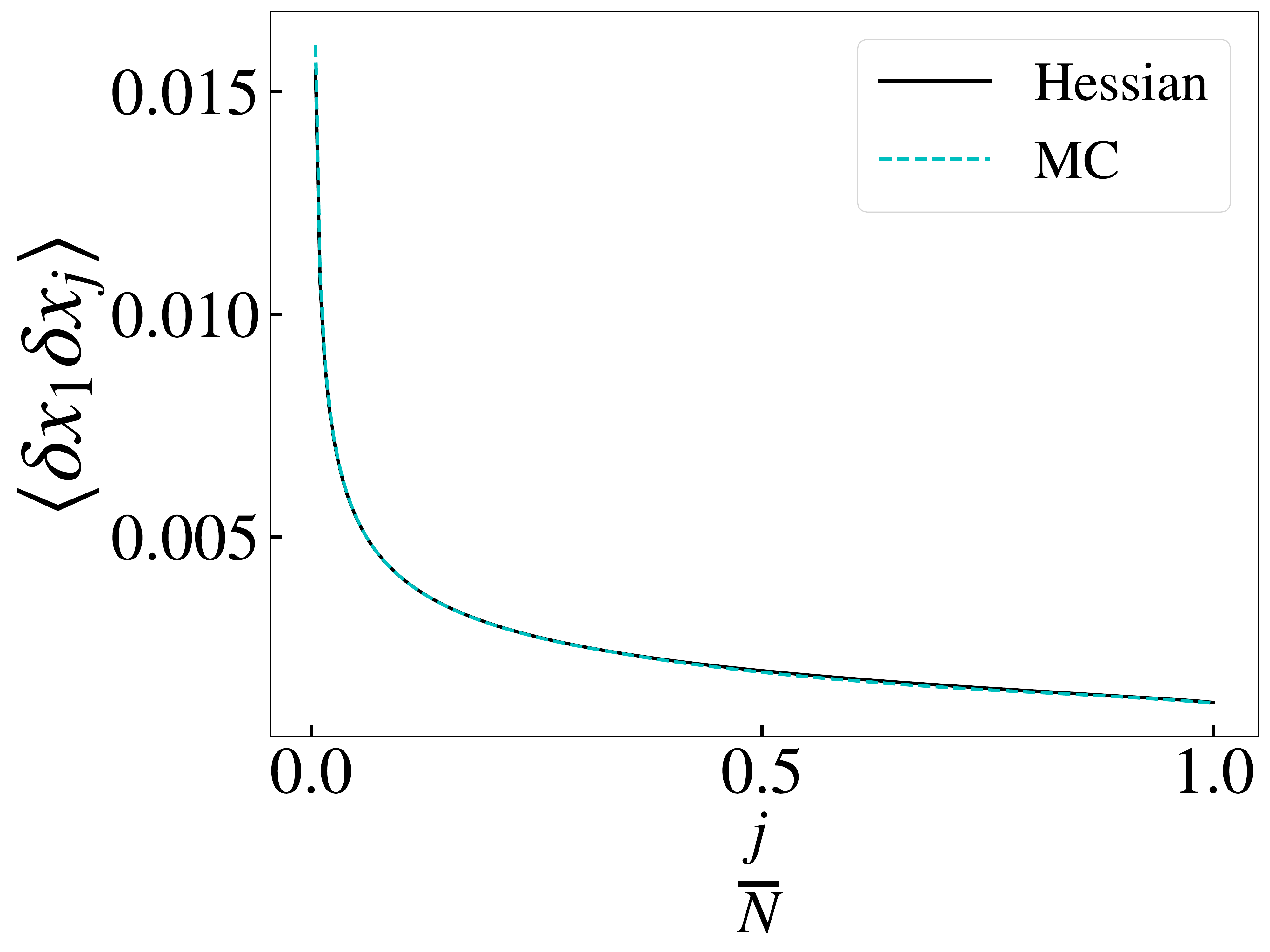}
        \caption{}
    \end{subfigure}
    \begin{subfigure}[b]{0.49\textwidth}
        \includegraphics[width=\linewidth]{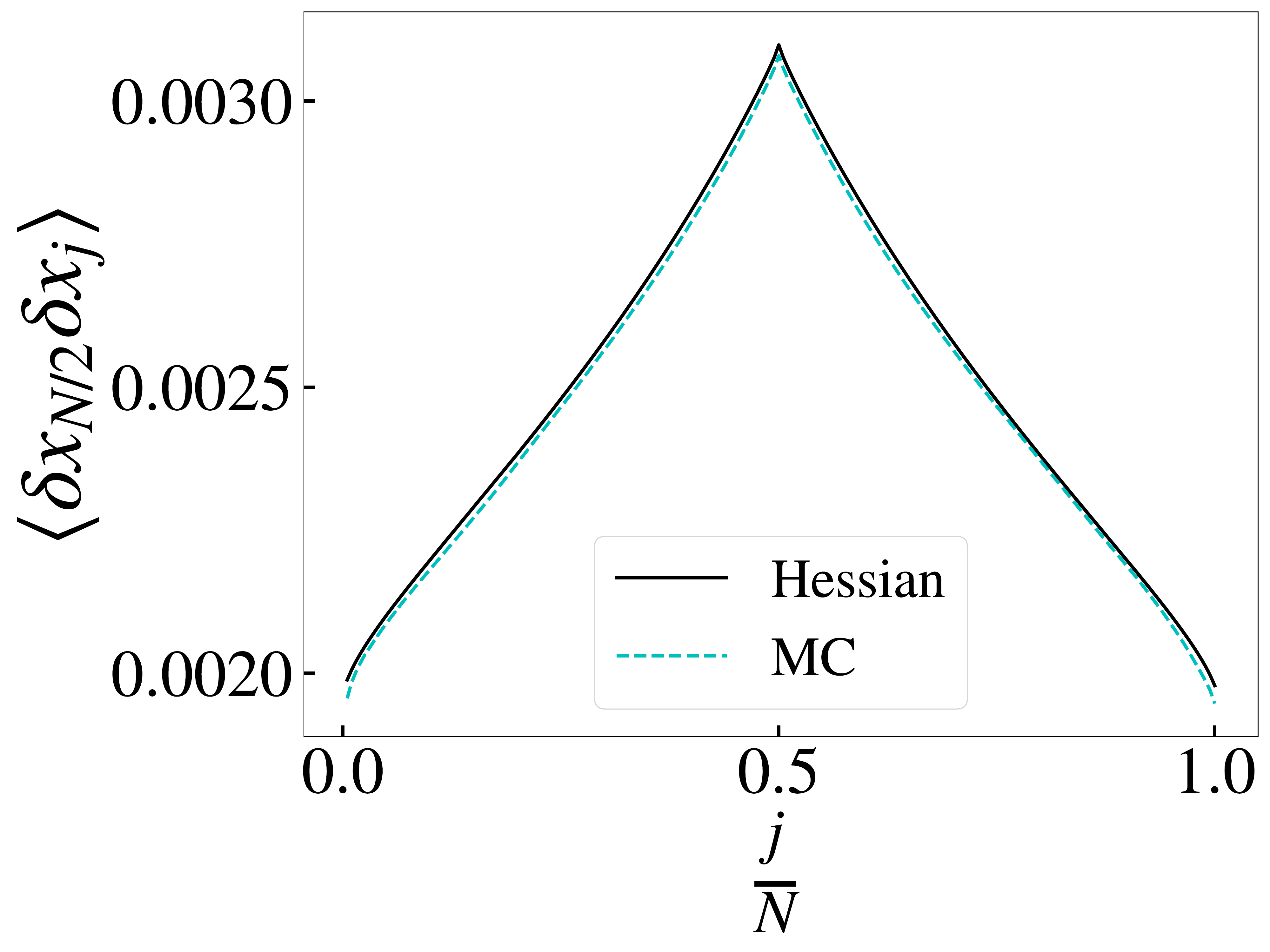}
        \caption{}
    \end{subfigure}
    \newline
    \begin{subfigure}[b]{0.49\textwidth}
        \includegraphics[width=\linewidth]{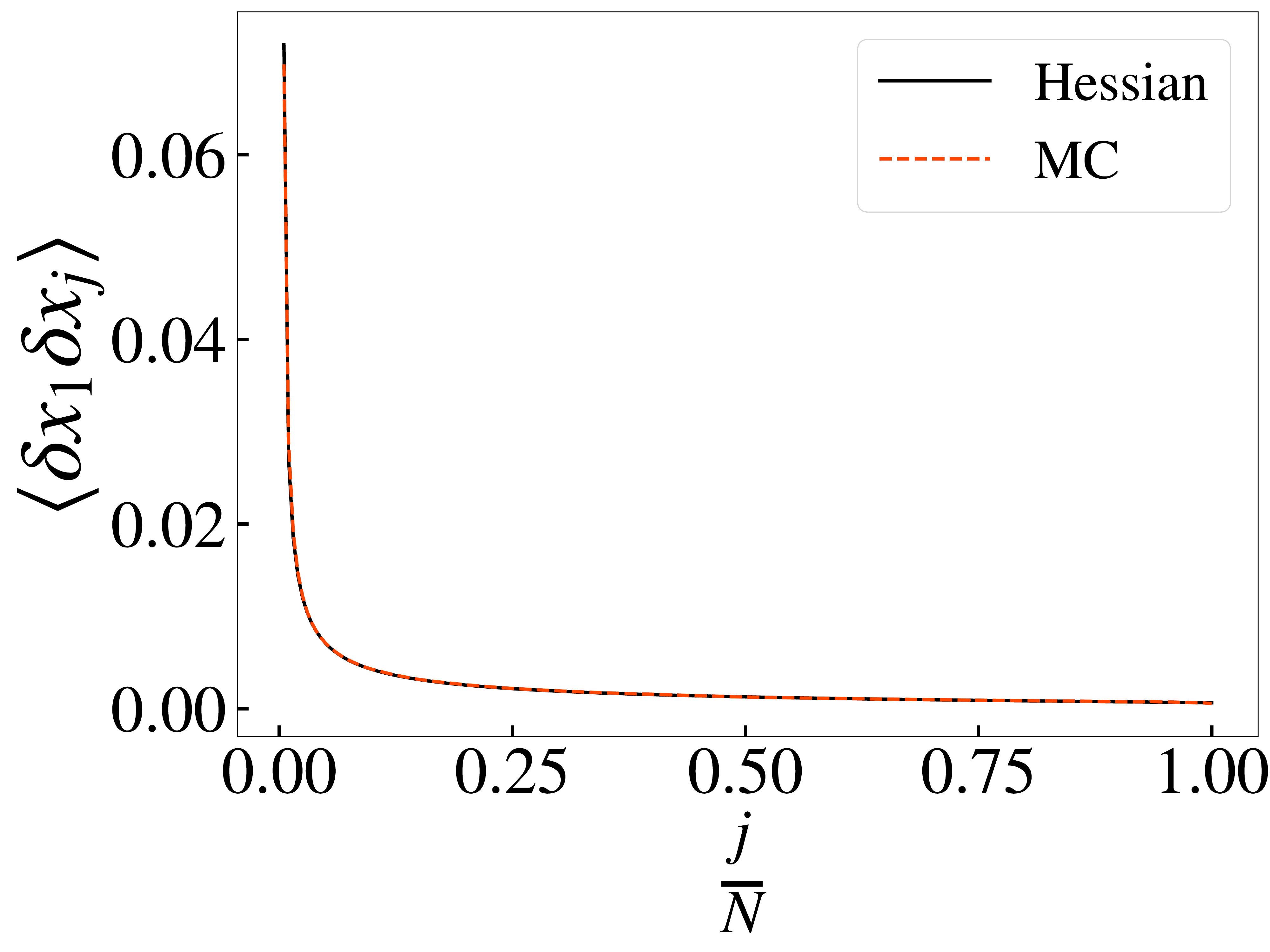}
        \caption{}
    \end{subfigure}
    \begin{subfigure}[b]{0.49\textwidth}
        \includegraphics[width=\linewidth]{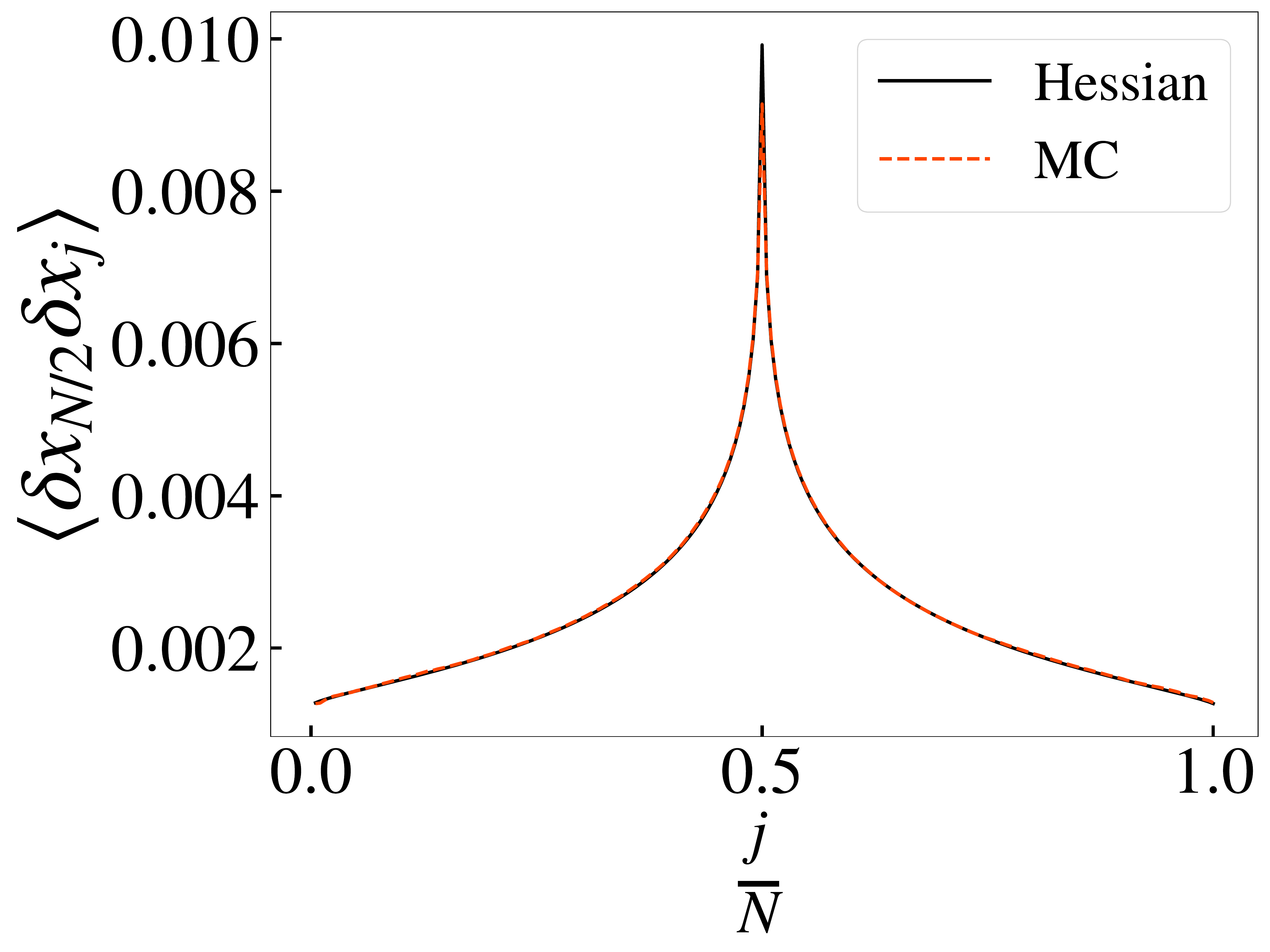}
        \caption{}
    \end{subfigure}
    \caption{Comparison of correlations  $\langle \delta x_1 \delta x_j \rangle$ and   $\langle \delta x_{N/2} \delta x_j \rangle$, obtained from MC simulations, with those obtained from the  Hessian matrix, for (a,b) CM model and (c,d) the LG model at $\beta=2$ and $N=200$. }
    \label{fig:C_ij for mid and last particles}
\end{figure}

Finally we present results on  correlations in the fluctuations in the positions of the ordered particles,  {\emph i.e.,} we look at $\langle \delta x_i \delta x_k \rangle $, where $\delta x_i = x_i -\langle x_i\rangle$. 
As stated in Eq.~(\ref{inv hessian element}), in the small oscillation approximation, the two-point correlations are given by the inverse of the Hessian matrix. In Figs.~(\ref{fig:C_ij for mid and last particles}) we compare results for  correlations obtained from direct numerical simulations with those obtained from the Hessian, for $N=200$. It is clear that there is  good agreement  for both the  CM and LG models. 

\begin{figure}[!h]
    \begin{subfigure}[b]{0.49\textwidth}
        \includegraphics[width=\linewidth]{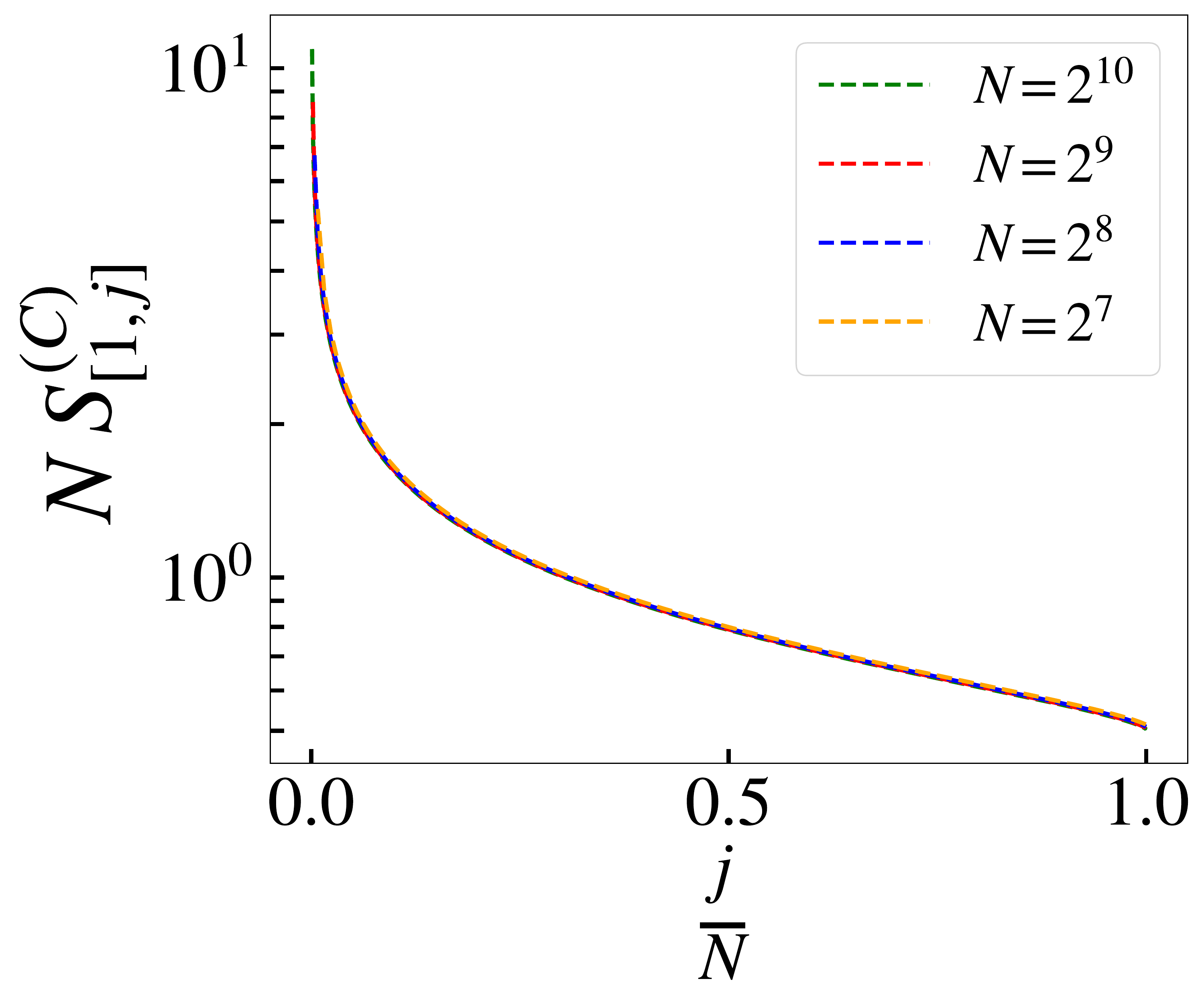}
        \caption{}
    \end{subfigure}
    \begin{subfigure}[b]{0.49\textwidth}
        \includegraphics[width=\linewidth]{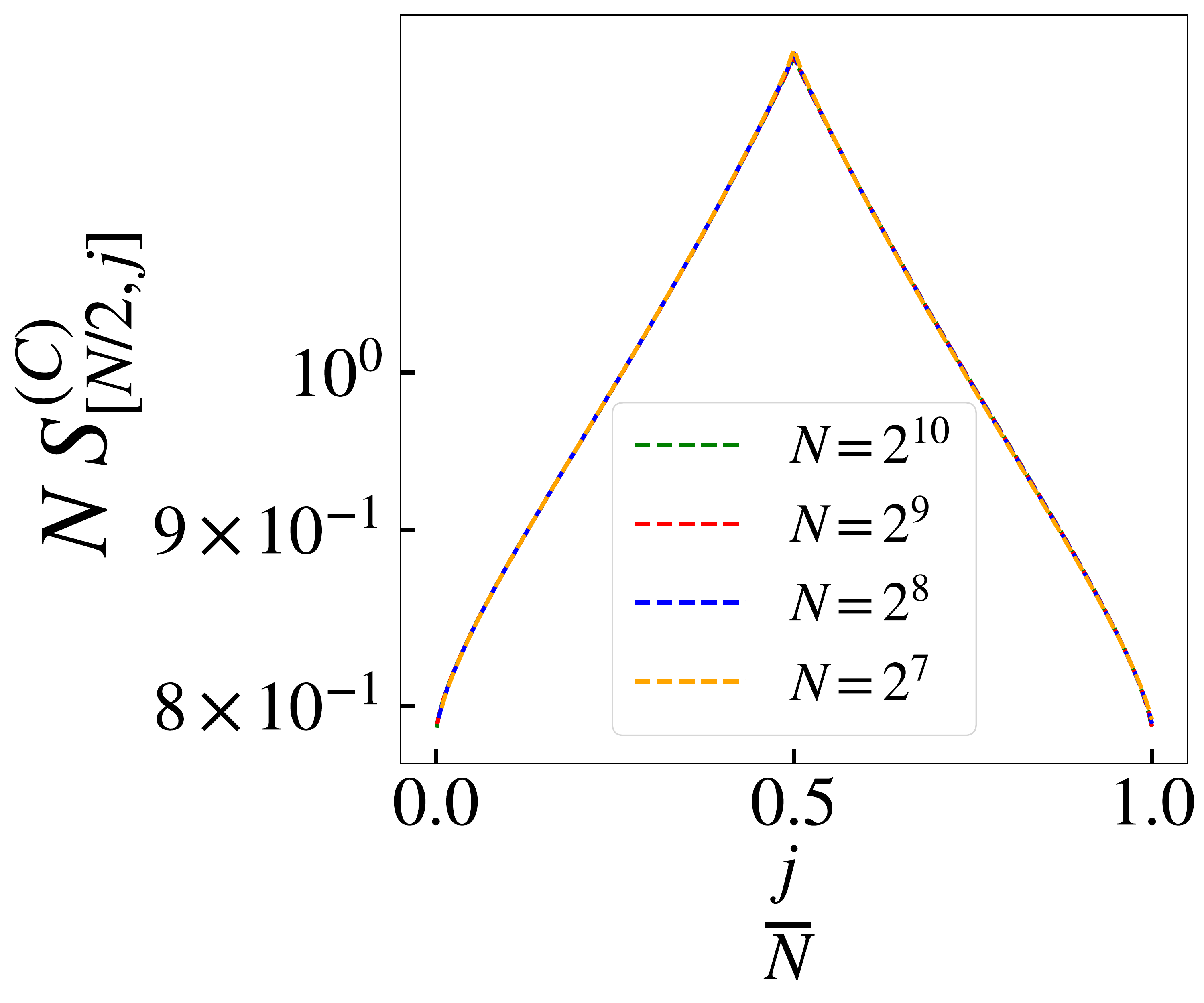}
        \caption{}
    \end{subfigure}
    \newline
    \begin{subfigure}[b]{0.49\textwidth}
        \includegraphics[width=\linewidth]{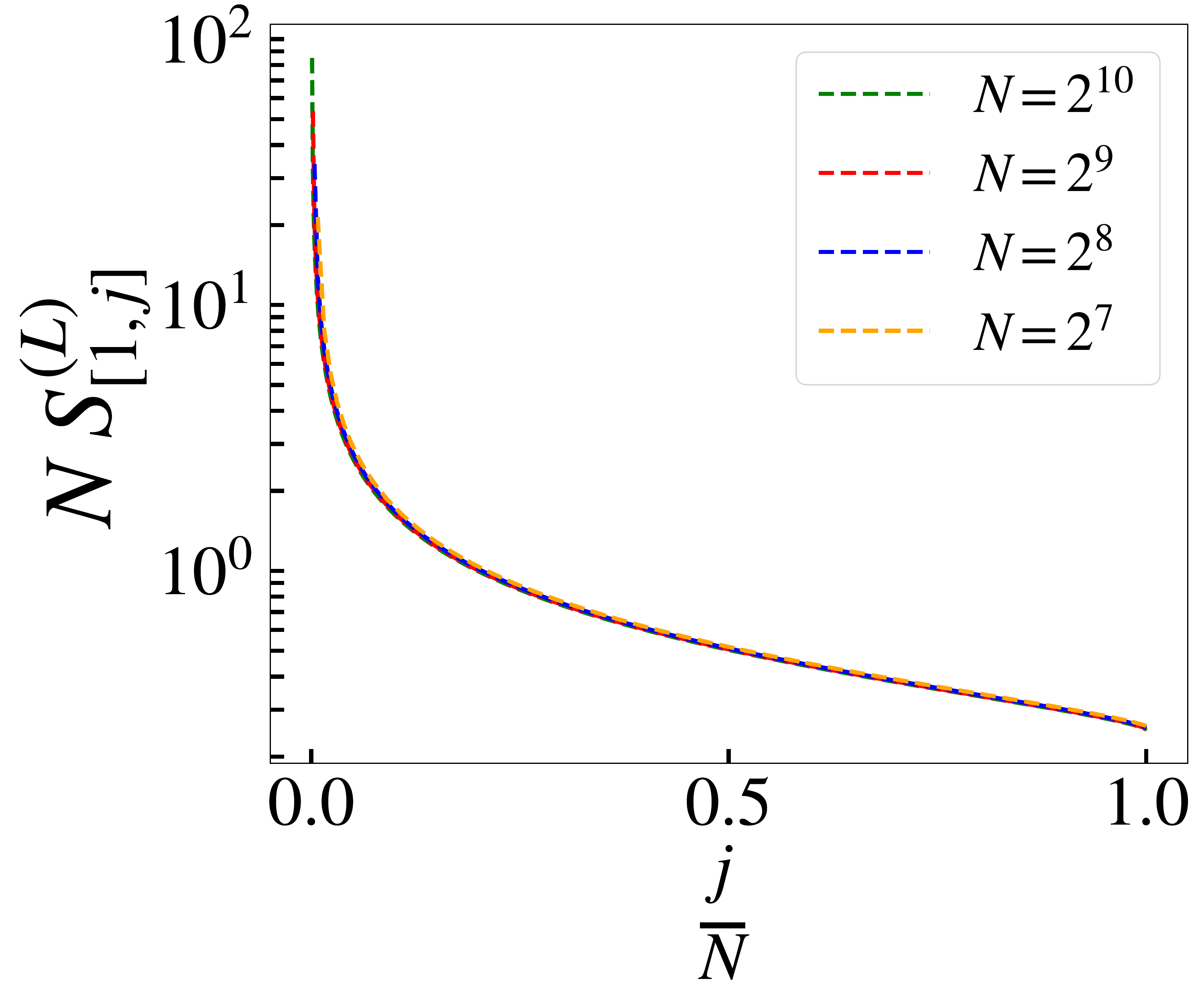}
        \caption{}
    \end{subfigure}
    \begin{subfigure}[b]{0.49\textwidth}
        \includegraphics[width=\linewidth]{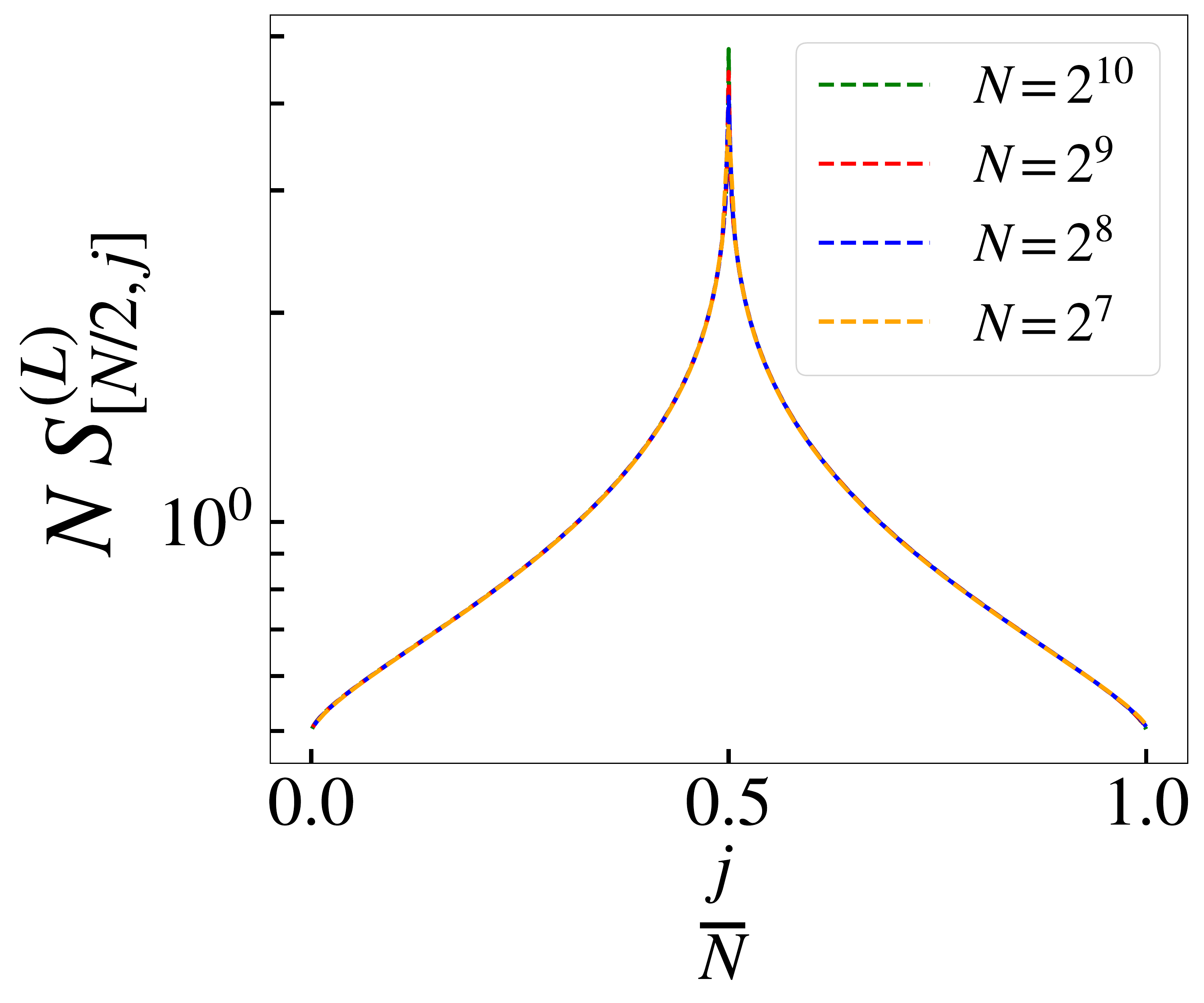}
        \caption{}
    \end{subfigure}
    \caption{Plot of correlations  $\langle \delta x_1 \delta x_j \rangle$ and  $\langle \delta x_{N/2} \delta x_j \rangle$ obtained from the Hessian matrix, for (a,b) the CM model and (c,d) the LG model. It is clear that the correlations satisfy a $1/N$ scaling} 
    \label{fig:N_scaling_inv_hess}
\end{figure}

In  Fig.~(\ref{fig:N_scaling_inv_hess}) we present results from the Hessian matrix for different system sizes 
and find the following finite-size scaling form for both models:
\begin{equation}
S^{(C/L)}_{[i,j]} = \frac{1}{N} f\left(\frac{i}{N}, \frac{j}{N}\right), 
\end{equation}
where $f$ is some scaling function. This has been shown  for both Calogero and Log-gas models. As shown in the figure, for Calogero model this relation holds well at all particle positions. Near the edge particles, the relation breaks down. For Log-gas, this relation holds reasonably well when correlations are between particles separated by $~O(N)$ but breaks down for smaller distances.

\section{Conclusions}
\label{conc}
In this work we pointed out some connections between the log-gas (LG) and the Calogero-Moser (CM) model, two systems that  have been widely studied in the context of random matrix theory and integrable models respectively. We looked at equilibrium properties of these two systems by performing extensive Monte-Carlo simulations to compute  single-particle distribution functions of edge and bulk particles, and two-point correlation functions. We compared the Monte-Carlo results with those obtained from a Hessian theory, corresponding to making  the approximation of small oscillations about the potential minima. We find that, except for the form of the distribution of the edge particle (which is non-Gaussian),  the small oscillation approximation results are in general found to be in close agreement with the Monte-Carlo results. This  includes results on non-trivial system-size scaling properties. 

For the LG model, our Monte-Carlo simulations provide an accurate verification of the Tracy-Widom distribution at the parameter values $\beta=1,2,4$. 
For the CM model, we find a non-trivial scaling of the MSD for the edge particle, $\langle \delta x_N^2 \rangle \sim  N^{-2/3}$, as opposed to $\langle \delta x_N^2 \rangle \sim  N^{-1/3}$ for the LG. However, while we find that the distribution is non-Gaussian, it is distinct from the Tracy-Widom form seen in the LG.  
For the LG model, our MC simulations and  Hessian theory computations results  
provide a verification of  recent predictions \cite{gustavsson2005gaussian,o2010gaussian,zhang2015gaussian} on single-particle  distribution of bulk particles, including the  $\log(N)/N$ scaling of the MSD.  For the CM model we find that fluctuations of bulk particles are Gaussian, with a $1/N$ scaling of the MSD, and two-point correlations also have the same scaling. Some further interesting future directions would be the analytical computation of correlations from the Hessian theory for both models and the use of field theory approaches for the CM model to obtain analytical results. 

\section{Acknowledgements}
We thank Anirban Basak, Manjunath Krishnapur, Anupam Kundu, Arul Lakshminarayan, Joseph Samuel, Herbert Spohn,   Patrik Ferrari and  Satya Majumdar for useful discussions. AD would like to acknowledge support from the project 5604-2 of the Indo-French Centre for the Promotion of Advanced Research (IFCPAR). MK would like to acknowledge support from the project 6004-1 of the Indo-French Centre for the Promotion of Advanced Research (IFCPAR). MK gratefully acknowledges the Ramanujan Fellowship SB/S2/RJN-114/2016 from the Science and Engineering Research Board (SERB), Department of Science and Technology, Government of India. SA would like to acknowledge support from the Long Term Visiting Students' Programme, 2018 at ICTS. SA is also grateful to Aditya Vijaykumar and Junaid Bhatt for helpful discussions. We would like to thank the ICTS program ``Universality in random structures: Interfaces, Matrices, Sandpiles  (Code: ICTS/URS2019/01)" for enabling valuable discussions with many participants. 

\section{Equilibration and Convergence check}

The most general form of the equipartition theorem states that under suitable assumptions, for a physical system with Hamiltonian $H$ and degrees of freedom $x_n$, the following holds in thermal equilibrium for all indices $m$ and $n$ \cite{pathria1986statistical}:
\begin{equation}
\left\langle x_m \frac{\partial H}{\partial x_n} \right\rangle = \delta_{mn} k_B T
\label{virial_eqn}
\end{equation}
This has been used to ascertain whether the system is at its equilibrium configuration, where $x_i$ is the position of $i^{th}$ particle. Using ergodicity, time average can be considered equivalent to ensemble average and hence, the above average is over microstates in the ensemble. Convergence check has been performed over different number of microstates (samples) to see if the number of samples is sufficient for averaging in statistics. We have found that the Calogero-Moser system converges at $16\times 10^7$ samples for $N = 100, 200$ and the Log-gas System converges at $8\times 10^7$ samples for $N = 100, 200$ as shown in Figure (\ref{fig:virial}). This figure demonstrates that our brute-force numerics are very accurate given the good agreement with the generalized virial theorem, which quantifies the equilibration of the system as given in equation (\ref{virial_eqn}). 
\afterpage{
\begin{figure}[!h]
    \begin{subfigure}[b]{0.47\textwidth}
        \includegraphics[width=\linewidth]{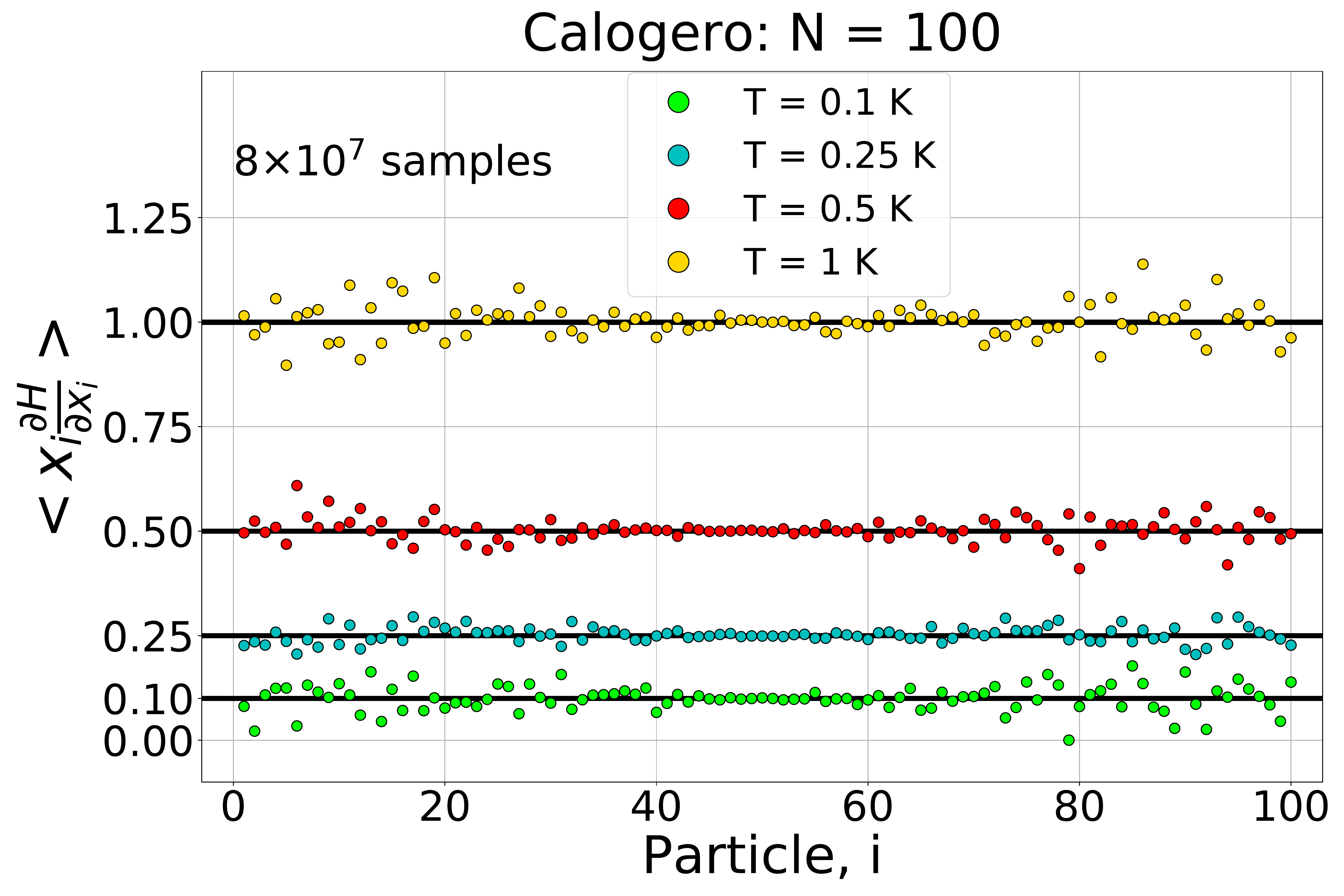}
        \caption{}
    \end{subfigure}
    \hfill
    \begin{subfigure}[b]{0.47\textwidth}
        \includegraphics[width=\linewidth]{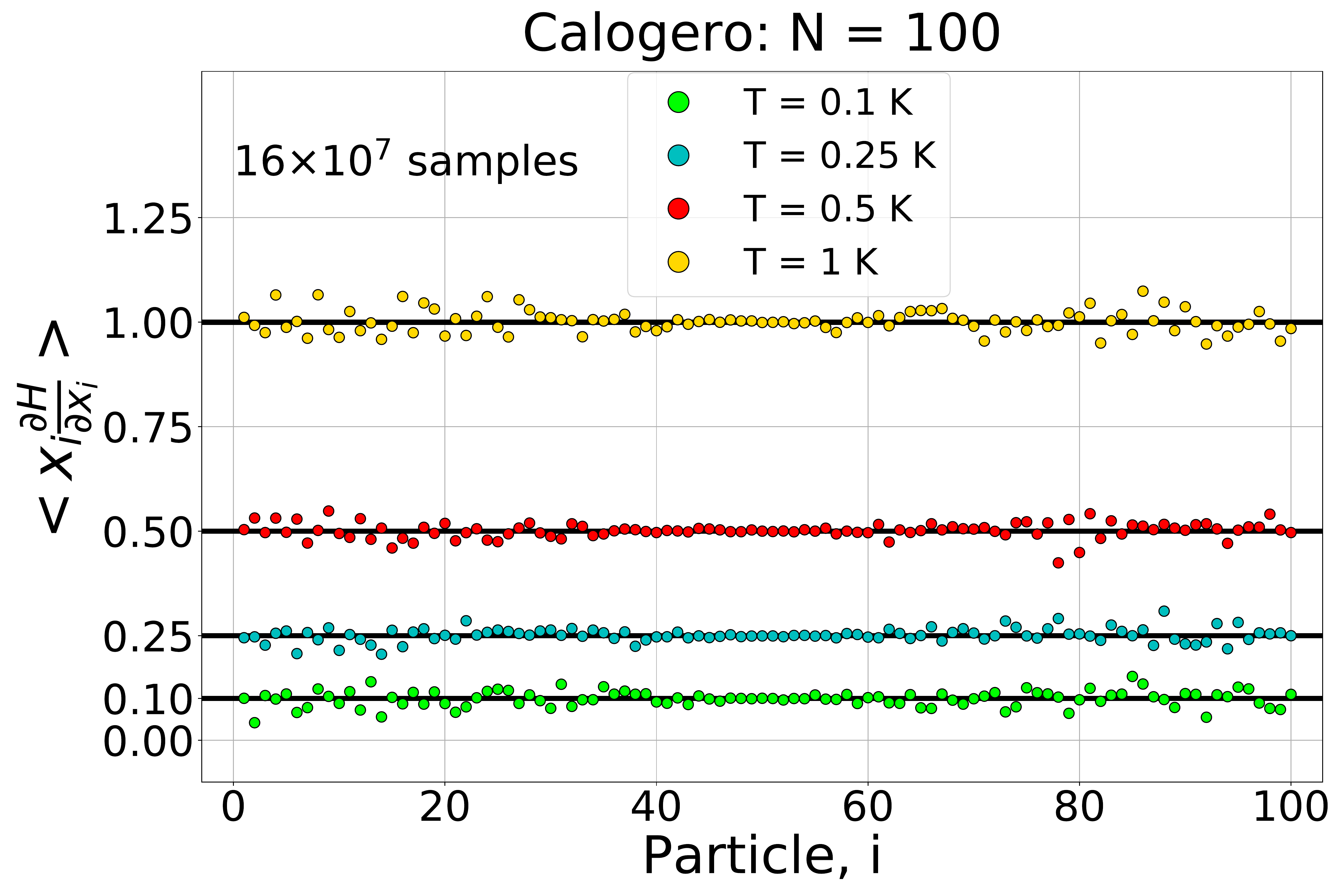}
        \caption{}
    \end{subfigure}
    \newline
    \begin{subfigure}[b]{0.47\textwidth}
        \includegraphics[width=\linewidth]{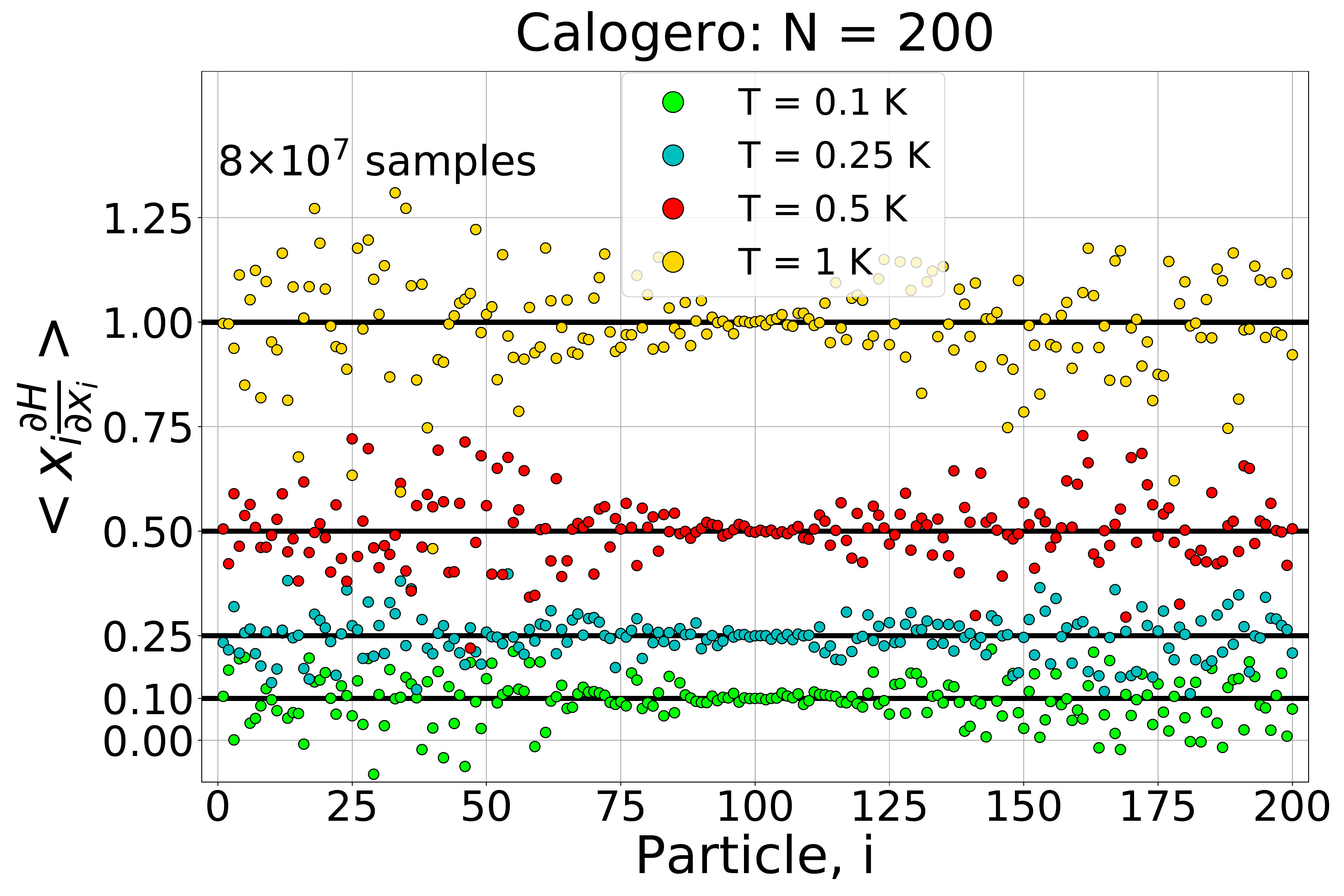}
        \caption{}
    \end{subfigure}
    \hfill
    \begin{subfigure}[b]{0.47\textwidth}
        \includegraphics[width=\linewidth]{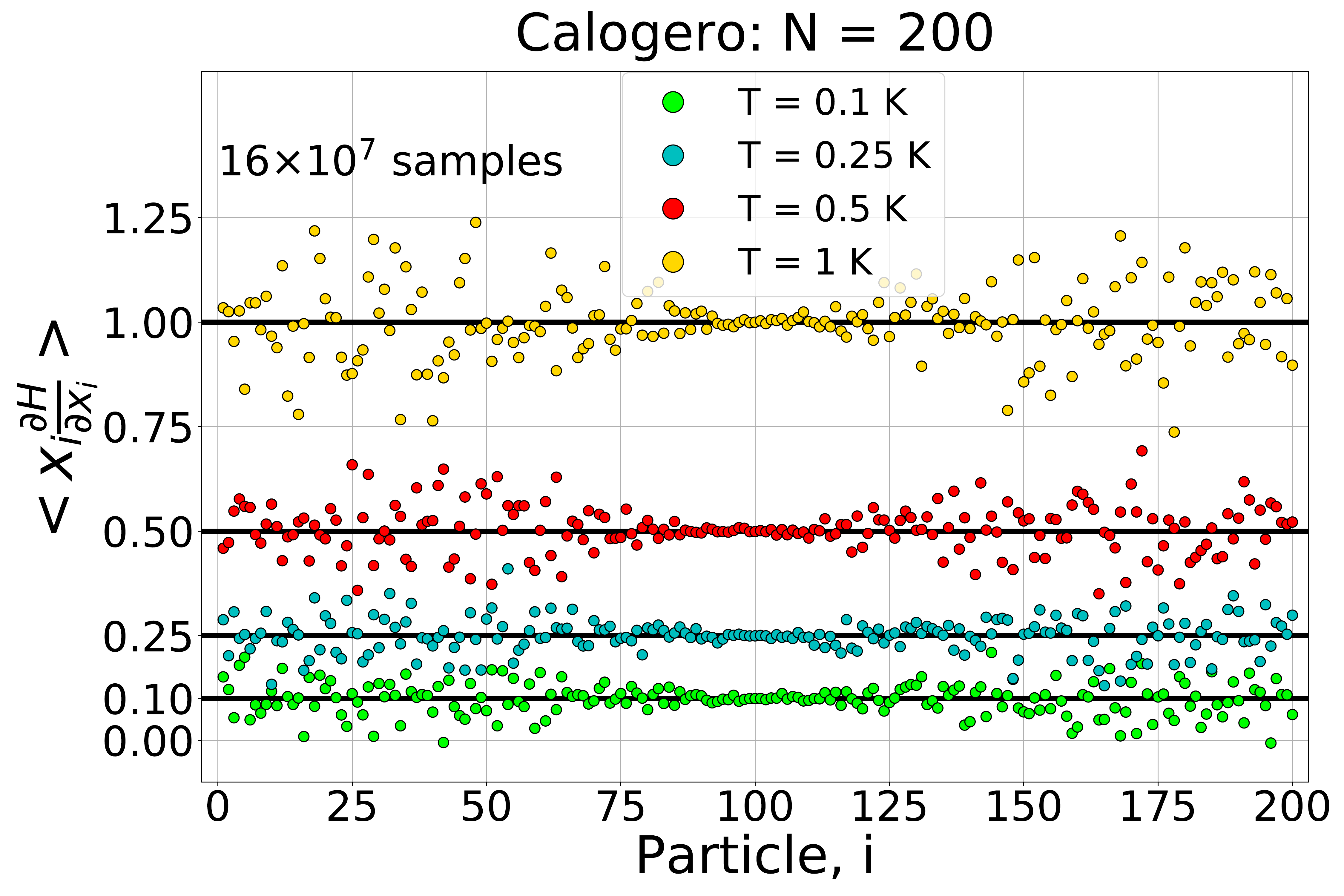}
        \caption{}
    \end{subfigure}
    \newline
    \begin{subfigure}[b]{0.47\textwidth}
        \includegraphics[width=\linewidth]{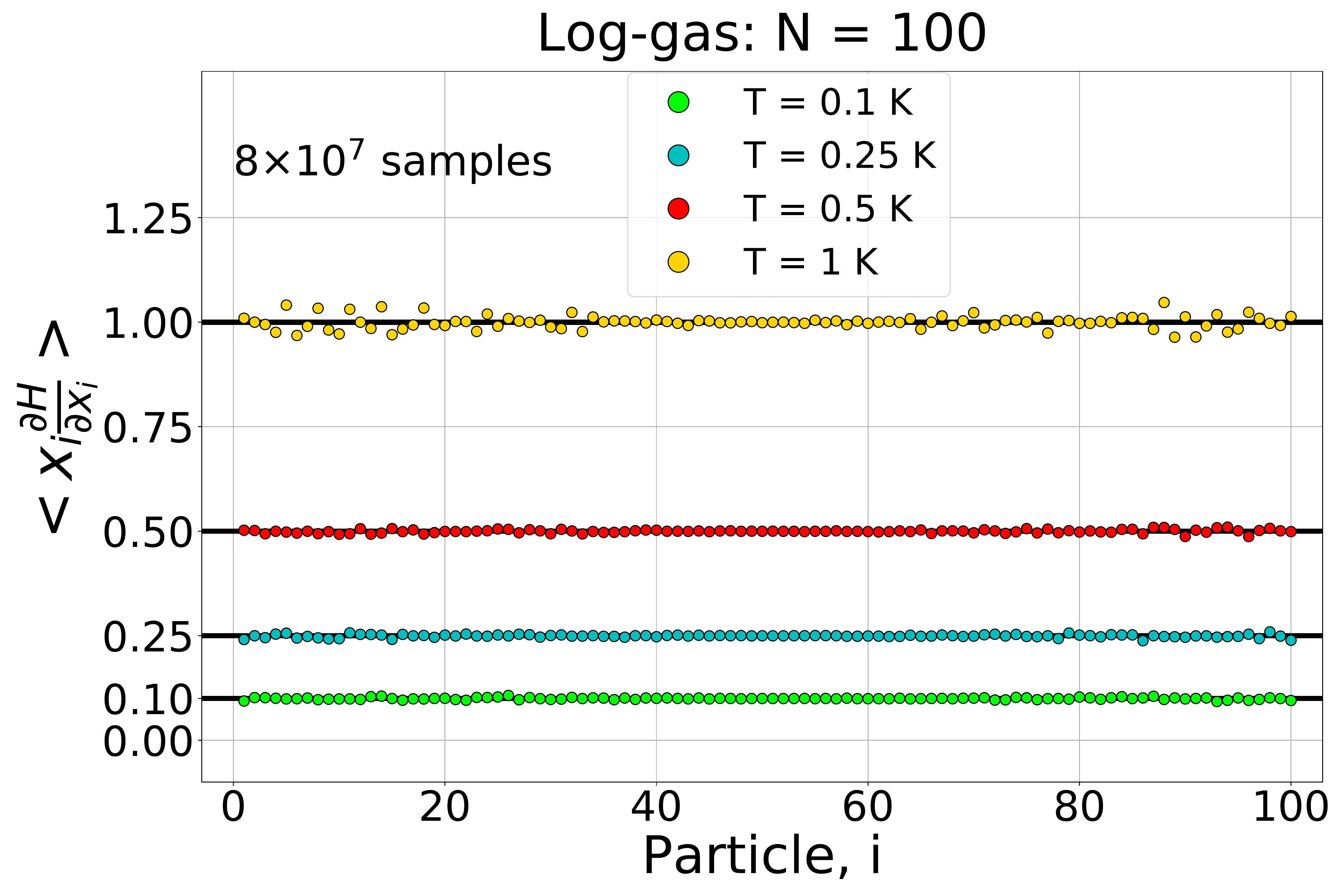}
        \caption{}
    \end{subfigure}
    \hfill
    \begin{subfigure}[b]{0.47\textwidth}
        \includegraphics[width=\linewidth]{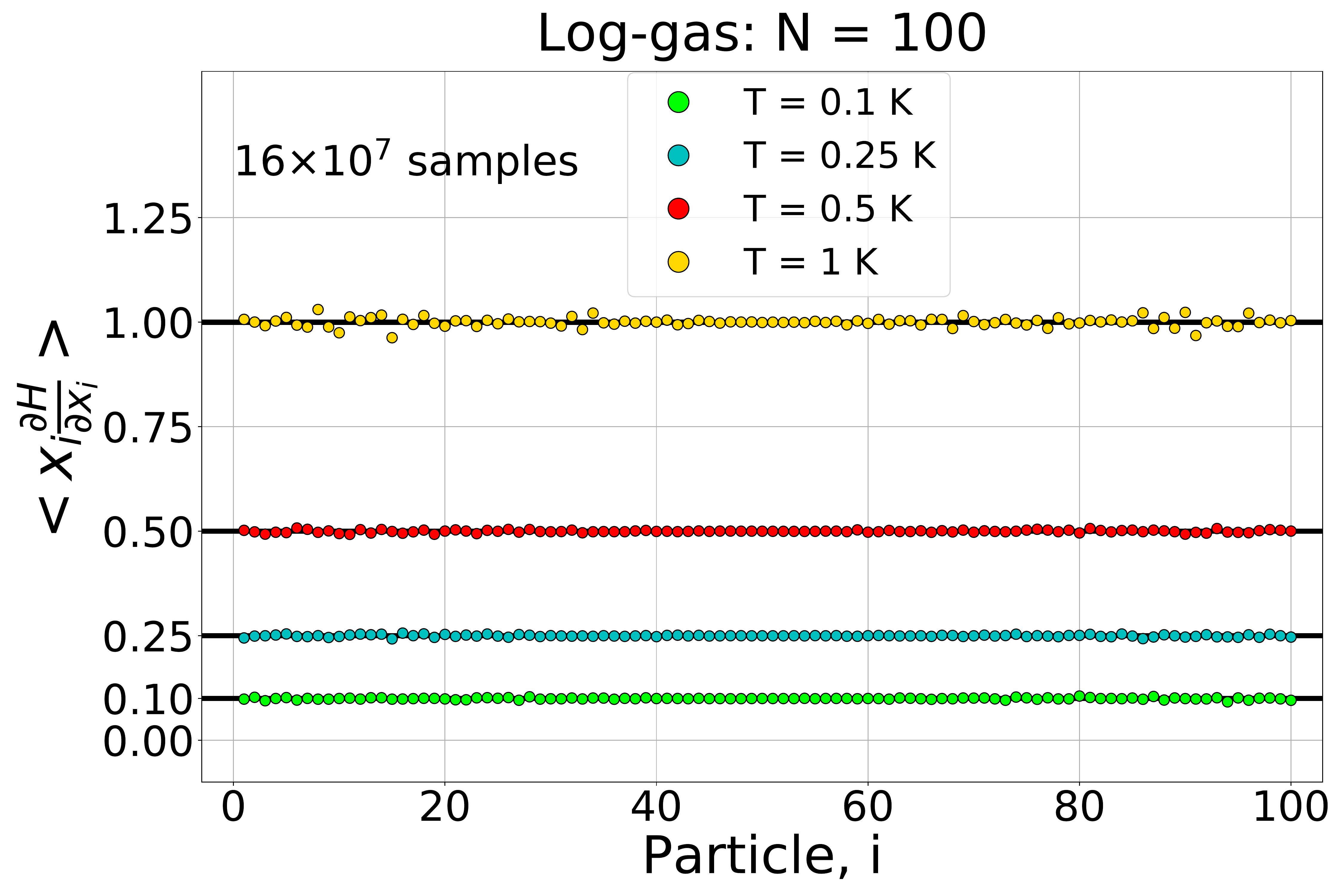}
        \caption{}
    \end{subfigure}
    \newline
        \begin{subfigure}[b]{0.47\textwidth}
        \includegraphics[width=\linewidth]{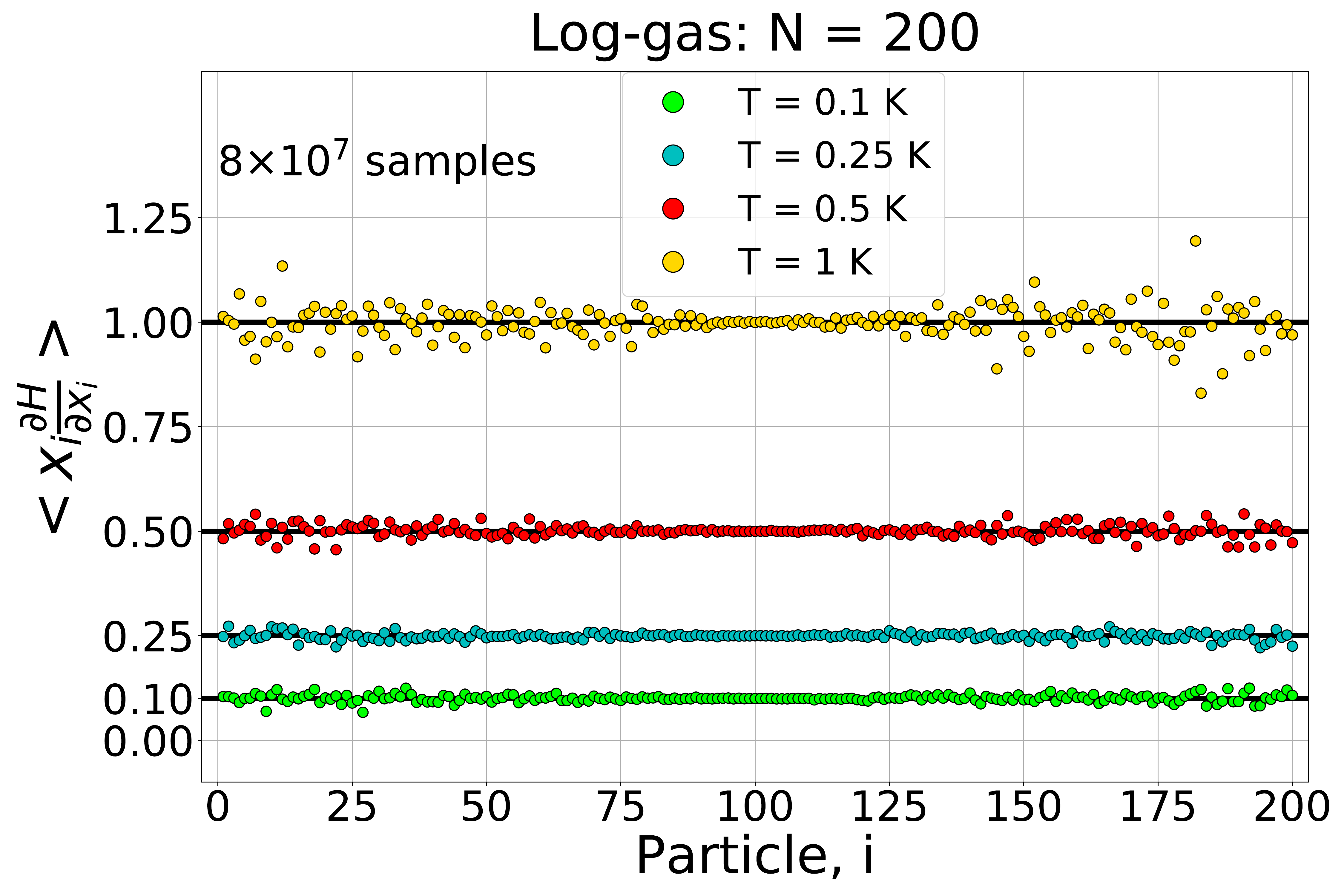}
        \caption{}
    \end{subfigure}
    \hfill
    \begin{subfigure}[b]{0.47\textwidth}
        \includegraphics[width=\linewidth]{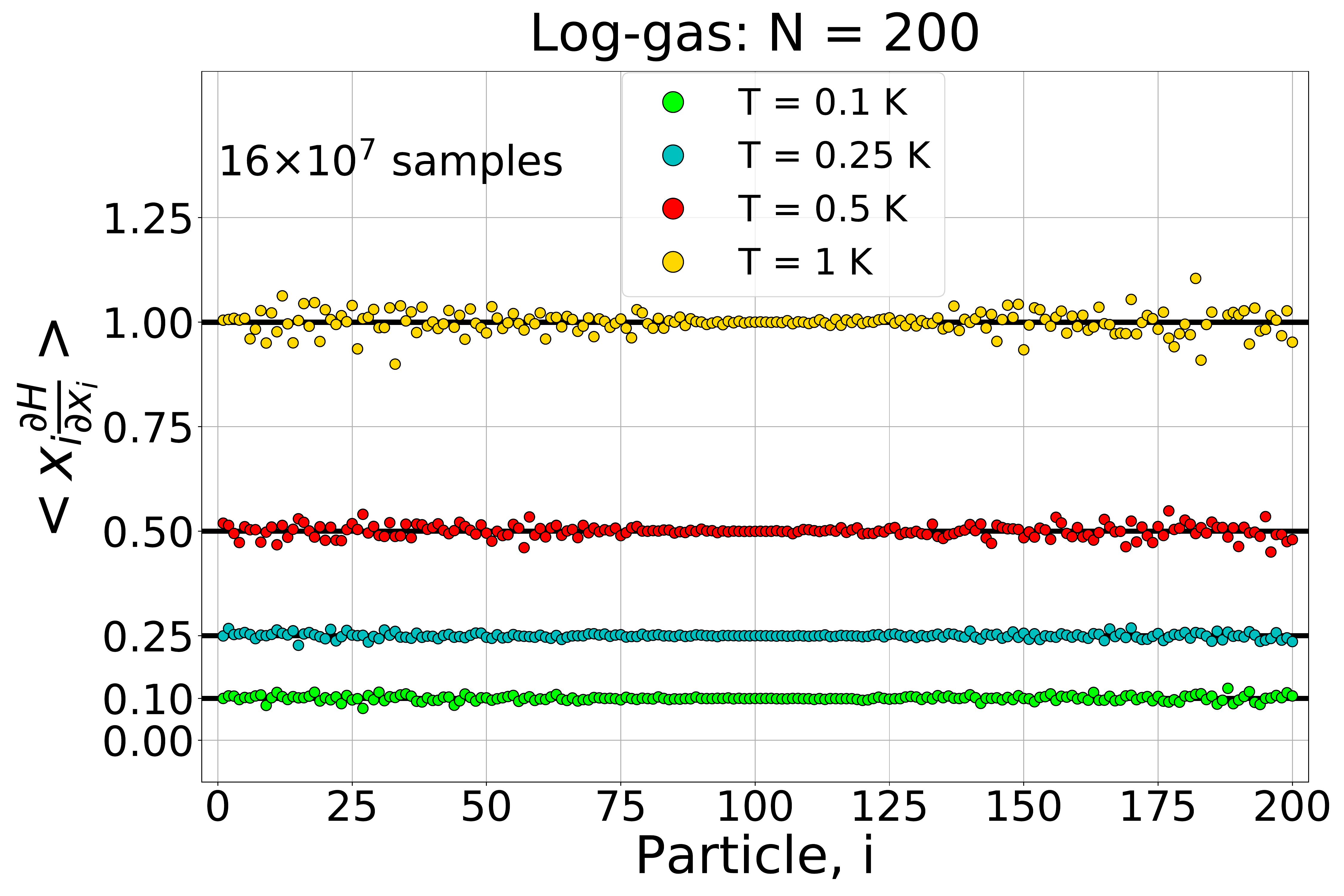}
        \caption{}
    \end{subfigure}
    \caption{The check for thermal equilibrium using equation (\ref{virial_eqn}) has been performed for both systems and we observe convergence with respect to sample size.}
    \label{fig:virial}
\end{figure}
\clearpage
}


\appendix
\label{appendix}

\newpage
\bibliographystyle{unsrt}
\bibliography{references}

\begin{thebibliography}{10}

\bibitem{forrester2010log}
Peter~J Forrester.
\newblock {\em Log-gases and random matrices (LMS-34)}.
\newblock Princeton University Press, 2010.

\bibitem{nagao1995asymptotic}
Taro Nagao and Peter~J Forrester.
\newblock Asymptotic correlations at the spectrum edge of random matrices.
\newblock {\em Nuclear Physics B}, 435(3):401--420, 1995.

\bibitem{bourgade2014bulk}
Paul Bourgade.
\newblock Bulk universality for one-dimensional log-gases.
\newblock In {\em XVIIth International Congress on Mathematical Physics}, pages
  404--416. World Scientific, 2014.

\bibitem{erdos2012universality}
L{\'a}szl{\'o} Erdos.
\newblock Universality for random matrices and log-gases.
\newblock {\em arXiv preprint arXiv:1212.0839}, 2012.

\bibitem{ameur2011fluctuations}
Yacin Ameur, H{\aa}kan Hedenmalm, Nikolai Makarov, et~al.
\newblock Fluctuations of eigenvalues of random normal matrices.
\newblock {\em Duke mathematical journal}, 159(1):31--81, 2011.

\bibitem{deift2006universality}
Percy Deift.
\newblock Universality for mathematical and physical systems.
\newblock {\em arXiv preprint math-ph/0603038}, 2006.

\bibitem{tracy1998correlation}
Craig~A Tracy and Harold Widom.
\newblock Correlation functions, cluster functions, and spacing distributions
  for random matrices.
\newblock {\em Journal of statistical physics}, 92(5-6):809--835, 1998.

\bibitem{tracy2009distributions}
Craig~A Tracy and Harold Widom.
\newblock The distributions of random matrix theory and their applications.
\newblock In {\em New trends in mathematical physics}, pages 753--765.
  Springer, 2009.

\bibitem{widom1999relation}
Harold Widom.
\newblock On the relation between orthogonal, symplectic and unitary matrix
  ensembles.
\newblock {\em Journal of Statistical Physics}, 94(3-4):347--363, 1999.

\bibitem{baik2005phase}
Jinho Baik, G{\'e}rard~Ben Arous, Sandrine P{\'e}ch{\'e}, et~al.
\newblock Phase transition of the largest eigenvalue for nonnull complex sample
  covariance matrices.
\newblock {\em The Annals of Probability}, 33(5):1643--1697, 2005.

\bibitem{baker1997finite}
TH~Baker and PJ~Forrester.
\newblock Finite-n fluctuation formulas for random matrices.
\newblock {\em Journal of statistical physics}, 88(5-6):1371--1386, 1997.

\bibitem{nagao1998transitive}
Taro Nagao and Peter~J Forrester.
\newblock Transitive ensembles of random matrices related to orthogonal
  polynomials.
\newblock {\em Nuclear Physics B}, 530(3):742--762, 1998.

\bibitem{mehta2004random}
Madan~Lal Mehta.
\newblock {\em Random matrices}, volume 142.
\newblock Elsevier, 2004.

\bibitem{gustavsson2005gaussian}
Jonas Gustavsson.
\newblock Gaussian fluctuations of eigenvalues in the gue.
\newblock In {\em Annales de l'Institut Henri Poincare (B) Probability and
  Statistics}, volume~41, pages 151--178. No longer published by Elsevier,
  2005.

\bibitem{o2010gaussian}
Sean O'Rourke.
\newblock Gaussian fluctuations of eigenvalues in wigner random matrices.
\newblock {\em Journal of Statistical Physics}, 138(6):1045--1066, 2010.

\bibitem{zhang2015gaussian}
Deng Zhang.
\newblock Gaussian fluctuations of eigenvalues in log-gas ensemble: Bulk case
  i.
\newblock {\em Acta Mathematica Sinica, English Series}, 31(9):1487--1500,
  2015.

\bibitem{bornemann2009numerical}
Folkmar Bornemann.
\newblock On the numerical evaluation of distributions in random matrix theory:
  a review.
\newblock {\em arXiv preprint arXiv:0904.1581}, 2009.

\bibitem{calogero1975exactly}
F~Calogero.
\newblock Exactly solvable one-dimensional many-body problems.
\newblock {\em Lettere al Nuovo Cimento (1971-1985)}, 13(11):411--416, 1975.

\bibitem{calogero1969_1}
Francesco Calogero.
\newblock Solution of a three-body problem in one dimension.
\newblock {\em Journal of Mathematical Physics}, 10(12):2191--2196, 1969.

\bibitem{calogero1971solution}
Francesco Calogero.
\newblock Solution of the one-dimensional n-body problems with quadratic and/or
  inversely quadratic pair potentials.
\newblock {\em Journal of Mathematical Physics}, 12(3):419--436, 1971.

\bibitem{moser1976three}
JURGEN MOSER.
\newblock Three integrable hamiltonian systems connected with isospectral
  deformations.
\newblock In {\em Surveys in Applied Mathematics}, pages 235--258. Elsevier,
  1976.

\bibitem{bogomolny2009random}
E~Bogomolny, Olivier Giraud, and C~Schmit.
\newblock Random matrix ensembles associated with lax matrices.
\newblock {\em Physical review letters}, 103(5):054103, 2009.

\bibitem{kulkarni2017emergence}
Manas Kulkarni and Alexios Polychronakos.
\newblock Emergence of the calogero family of models in external potentials:
  duality, solitons and hydrodynamics.
\newblock {\em Journal of Physics A: Mathematical and Theoretical},
  50(45):455202, 2017.

\bibitem{polychronakos2006physics}
Alexios~P Polychronakos.
\newblock The physics and mathematics of calogero particles.
\newblock {\em Journal of Physics A: Mathematical and General}, 39(41):12793,
  2006.

\bibitem{olshanetsky1981classical}
MA~Olshanetsky and Askolʹd~Mikha{\u\i}lovich Perelomov.
\newblock Classical integrable finite-dimensional systems related to lie
  algebras.
\newblock {\em Physics Reports}, 71(5):313--400, 1981.

\bibitem{perelomov1990integrable}
Askold~Mikhailovich Perelomov.
\newblock {\em Integrable systems of classical mechanics and Lie algebras}.
\newblock Birkh{\"a}user, 1990.

\bibitem{abanov2011soliton}
Alexander~G Abanov, Andrey Gromov, and Manas Kulkarni.
\newblock Soliton solutions of a calogero model in a harmonic potential.
\newblock {\em Journal of Physics A: Mathematical and Theoretical},
  44(29):295203, 2011.

\bibitem{aniceto2010poisson}
In{\^e}s Aniceto, Jean Avan, and Antal Jevicki.
\newblock Poisson structures of calogero--moser and ruijsenaars--schneider
  models.
\newblock {\em Journal of Physics A: Mathematical and Theoretical},
  43(18):185201, 2010.

\bibitem{stone}
Inaki~Anduaga Michael~Stone and Lei Xing.
\newblock The classical hydrodynamics of the calogero–sutherland model.
\newblock {\em Journalof Physics A:Mathematical and Theoretical}, 41, June
  2008.

\bibitem{franchini2015universal}
Fabio Franchini, Andrey Gromov, Manas Kulkarni, and Andrea Trombettoni.
\newblock Universal dynamics of a soliton after an interaction quench.
\newblock {\em Journal of Physics A: Mathematical and Theoretical},
  48(28):28FT01, 2015.

\bibitem{franchini2016hydrodynamics}
Fabio Franchini, Manas Kulkarni, and Andrea Trombettoni.
\newblock Hydrodynamics of local excitations after an interaction quench in 1d
  cold atomic gases.
\newblock {\em New Journal of Physics}, 18(11):115003, 2016.

\bibitem{calogero1977equilibrium}
F~Calogero.
\newblock Equilibrium configuration of the one-dimensionaln-body problem with
  quadratic and inversely quadratic pair potentials.
\newblock {\em Lettere al Nuovo Cimento (1971-1985)}, 20(7):251--253, 1977.

\bibitem{sutherland1971quantum}
Bill Sutherland.
\newblock Quantum many-body problem in one dimension: Ground state.
\newblock {\em Journal of Mathematical Physics}, 12(2):246--250, 1971.

\bibitem{sutherland1972exact}
Bill Sutherland.
\newblock Exact results for a quantum many-body problem in one dimension. ii.
\newblock {\em Physical Review A}, 5(3):1372, 1972.

\bibitem{dhar2017}
Abhishek Dhar, Anupam Kundu, Satya~N. Majumdar, Sanjib Sabhapandit, and
  Gr\'egory Schehr.
\newblock Exact extremal statistics in the classical 1d coulomb gas.
\newblock {\em Phys. Rev. Lett.}, 119:060601, 2017.

\bibitem{doi:10.1137/0517035}
P.~Forrester and J.~Rogers.
\newblock Electrostatics and the zeros of the classical polynomials.
\newblock {\em SIAM Journal on Mathematical Analysis}, 17(2):461--468, 1986.

\bibitem{calogero1981matrices}
F~Calogero.
\newblock Matrices, differential operators, and polynomials.
\newblock {\em Journal of Mathematical Physics}, 22(5):919--934, 1981.

\bibitem{wigner1951statistical}
Eugene~P Wigner.
\newblock On the statistical distribution of the widths and spacings of nuclear
  resonance levels.
\newblock In {\em Mathematical Proceedings of the Cambridge Philosophical
  Society}, volume~47, pages 790--798. Cambridge University Press, 1951.

\bibitem{nadal2011TWderiv}
Celine Nadal and Satya~N Majumdar.
\newblock A simple derivation of the tracy--widom distribution of the maximal
  eigenvalue of a gaussian unitary random matrix.
\newblock {\em Journal of Statistical Mechanics: Theory and Experiment},
  2011(04):P04001, 2011.

\bibitem{szeg1939orthogonal}
Gabor Szeg.
\newblock {\em Orthogonal polynomials}, volume~23.
\newblock American Mathematical Soc., 1939.

\bibitem{pathria1986statistical}
RK~Pathria.
\newblock Statistical mechanics, international series in natural philosophy,
  1986.

\end{thebibliography}
\end{document}